\documentclass{aa}  

\usepackage{graphicx}

\usepackage{multicol, multirow}
\usepackage{mathtools}
\usepackage{amsmath, amsfonts}
\usepackage{graphicx}
\usepackage{bm}
\usepackage{hhline}

\usepackage{adjustbox}
\usepackage{tikz}
\usetikzlibrary{positioning}
\usetikzlibrary{arrows}
\usetikzlibrary{shapes, backgrounds}

\usepackage{rotating}

\usepackage{xspace}
\usepackage{makecell}

\usepackage{placeins}

\usepackage{rotating}
\usepackage{array}
\newcolumntype{x}[1]{>{\centering\let\newline\\\arraybackslash\hspace{0pt}}p{#1}}

\usepackage{hyperref}
\usepackage{txfonts}

 \title{Chemical Evolution in Nuclear Stellar Discs}

   \author{Jennifer K. S. Friske
          \inst{1}
          \and
          Ralph Sch{\"o}nrich\inst{2}
          }

   \institute{Kapteyn Astronomical Institute, University of Groningen, Postbus 800, NL-9700 AV Groningen, The Netherlands\\
              \email{friske@astro.rug.nl}
         \and
         Mullard Space Science Laboratory, University College London, Holmbury St. Mary, Dorking, Surrey, RH5 6NT, UK\\
             \email{r.schoenrich@ucl.ac.uk}
             }

   \date{Received XXX; accepted XXX}

  \abstract
   {Nuclear Stellar Discs (NSDs) have been observed in the vast majority of barred disc galaxies including the Milky Way. Their intense star formation is sustained by the intense gas inflows driven by their surrounding bars and frequently supports a large-scale galactic fountain. Despite their central role in galaxy evolution, their chemical evolution remains largely unexplored.}
   {We argue that the chemical composition of NSDs is best understood relative to the bar tips from which their gas is drawn.
    We make predictions of the detailed abundance profiles of gas and young stars within the NSD under different accretion scenarios from the galactic bar.}
   {We present the first systematic, multizonal modelling of the chemical evolution of nuclear stellar discs based on the RAMICES II code.}
   {We show that due to their different star formation history to galactic discs, NSDs offer a unique laboratory to break parameter degeneracies in chemical evolution models.  This allows us to identify the effects of the main parameters guiding NSD nucleosynthesis and disentangle them from the global enrichment history. We also show for two edge cases how the mode of gas accretion onto the NSD imprints on the gas abundance profiles and make predictions that can be tested with observations.}
   {}


\begin{document}

   \maketitle

\section{Introduction}

This paper aims to connect chemical evolution modelling to the physics and history of galactic nuclear stellar discs (NSDs).
The nature of this work is explorative - there is only a couple of pioneering 1-zone chemical evolution models \citep{grieco_chemical_2015} available, and as we will show in the paper, there is to date no consensus in the available spectroscopic data. Hence, our task here is to outline chemical evolution models and to predict how different scenarios, specifically regarding the gas accretion of NSDs, translate into observables.

Galactic NSDs are commonly found in the centres of barred galaxies including our own Milky Way (MW). Early hints were found in photometric data \citep{catchpole_distribution_1990} and its rotation detected in OH/IR stars \citep{lindqvist_ohir_1992}. Later, \citet{launhardt_nuclear_2002} suggested that the "nuclear bulge" had a stellar nuclear disc component, while not being able to rule out other configurations (e.g. bar/bulge). The existence of a NSD was finally proven through stellar kinematics in APOGEE spectroscopic data by \citet{schonrich_kinematic_2015}, revealing comparably cold kinematics  and a denser ring at the edge of the disc that strongly contrast with the surrounding bulge.  Usually the molecular gas (like in the Milky Way) is concentrated in a ring near the edge of the disc, where consequently also the young bright stars are concentrated, giving rise to a frequent term "nuclear ring", which however, is just part of the NSD as a whole.\footnote{Even though the term NSD technically refers to the stellar component, for the sake of simplicity, we use it here to refer to the whole galactic component, i.e. specifically also to the gas bound in the NSD.} The principles of NSD formation are qualitatively well understood: galactic bars funnel large amounts of gas into the central regions of galaxies, where it is predominantly deposited onto intensely star-forming rings at the edges of NSDs. We will elaborate on this in Sec.~\ref{sec:Background}.

However, we know less about their structure and evolution: there is no clear picture on how the material funnelled onto NSDs is redistributed beyond the star-forming ring, i.e. if the funnelled gas flows radially inwards through the NSD or is otherwise directly deposited further in. 
We tackle this in this work by tracing two edge cases of gas accretion, predicting their abundance profile and showing how they can be constrained by better spectroscopic data in the future.

We also lack information on the loss of gas and stellar yields from NSDs, other accretion mechanisms, internal migration/redistribution of stars, impact by the nearby active galactic nuclei (AGNs), etc. 
This is a strength of a full chemical evolution framework: since their inception \citep{burbidge_synthesis_1957,alpher_origin_1948,arnett_explosive_1970,pagel_metal_1975}, chemical evolution models have been used to give an independent constraint on galactic evolution. For example, they showed that galaxies must follow an open rather than closed box scheme with consistent inflow of fresh material \citep{tinsley_stellar_1979} and. more recently, radial abundance gradients have been quantitatively connected to inside-out formation \citep{schonrich_understanding_2017} and the mode of gas accretion from the corona \citep{bilitewski_radial_2012}. Furthermore, stellar abundances, when combined with chemical evolution models, have been employed to measure the radial migration of stars. Detailed models have also been used to constrain cosmic infall \citep[e.g.][]{colavitti_chemical_2008}, stochastic chemical evolution in the halo \citep[e.g.][]{matteucci_stochastic_1983, karlsson_stochastic_2005, cescutti_inhomogeneous_2008}, or to play on the stellar initial mass function (IMF) \citep[e.g.][]{matteucci_metallicity_1990,matteucci_light_1999}.

To use this approach for the NSD, a full multi-zone chemical evolution model of the Galactic centre must be drawn up and compared to the data. However, apart from a pioneering one-zone model by \citet{grieco_chemical_2015} \citep[also used in][]{thorsbro_detailed_2020} which covered some of the important aspects, no such work appears to have been undertaken. We will also make the point that our models are inherently predictions, because no sufficient data has been gathered on NSD abundances. 
Once data is available, a comparison will have powerful implications. For example, the overall metal content of an NSD can directly quantify the loss of yields and gas from it, by this determining (together with the total stellar mass) the total amount of gas that has been funnelled by the bar, and by this the angular momentum transfer to the bar. This angular momentum transfer has various important consequences and is for example essential to quantifying the angular momentum loss of the (slowing) bar to the dark halo and by this quantifying the structure of the dark halo \citep{chiba_resonance_2020,chiba_tree-ring_2021}. The mass of the NSD alone hereby only provides an upper bound on the total inflow integrated over time and cannot account for the currently unconstrained outflows. Therefore, NSD chemistry will provide crucial information for dynamic models of the Milky Way.

Further than the study of NSD physics, chemical evolution models are in dire need of different laboratories than galactic discs to break weakly constrained parameter degeneracies in chemical evolution. As one important example, it has been noted that the only way to resolve the chemical evolution of r-process elements with significantly super-solar $\left[{\rm r-process}/\alpha\right]$ ratios \citep{cote_origin_2018,cote_neutron_2019} is using a two--phase interstellar medium (ISM) and different contributions to the cold star forming gas phase for neutron star mergers  and core collapse supernovae  (\citealp{schonrich_chemical_2019, fraser_metallicity-suppressed_2021}, see also Sec.~\ref{sec:Background}). However, this hot/cold gas phase separation introduces a whole set of new parameters, that thus far are only constrained by this single observation, as chemical evolution becomes insensitive to this parameter at timescales larger than the hot gas cooling timescale ($\sim 1$ Gyr). Here the NSD can come to the rescue: due to its particular flow pattern and large expected loss rates of hot gas, it can display a permanent difference in abundance ratios, thus breaking this parameter degeneracy.

The paper is structured as follows:
We will first give an overview of the necessary theoretical background about chemical evolution and the current state of the research about NSDs in section~\ref{sec:Background}. 
Then, we introduce the models used and the adjustments made to incorporate an NSD in section~\ref{sec:Model}. Section~\ref{sec:Results} then examines how the distinct star formation history of the NSD shapes its gas abundance profiles. Furthermore, due to the largely unknown method of gas accretion onto the nuclear stellar disc, we explore the effects of two different gas accretion scenarios. Finally, section~\ref{sec:Conclusion} provides a brief overview of future research directions.

\section{Background}
\label{sec:Background}

\subsection{Elemental Abundances} 
Chemical evolution models connect physical parameters of a system to the chemical composition of ISM and stars, resolved by age and position. The star formation history, accretion, etc. in a galaxy affect the enrichment of the star-forming ISM with processed gas from previous generations. However, stellar age determination is difficult and fraught with uncertainties, so \cite{tinsley_stellar_1979} pioneered the use of elemental abundance ratios as clocks instead:
\begin{equation*}
{\rm [X/Fe]} = \log_{10}\frac{n_{\rm X}}{n_{\rm Fe}} - \log_{10}\frac{n_{\rm X, \odot}}{n_{\rm Fe,\odot}}.
\end{equation*}
Hereby, $n_X$ denotes the absolute or relative abundance of element $X$ and $n_{X,\odot}$ the corresponding solar abundance. The latter is somewhat uncertain; here we use the abundances from \citet{magg_observational_2022}.

\subsection{Timescales in Chemical Evolution} 
\begin{figure}
    \centering
    \begin{adjustbox}{max totalsize={.99\columnwidth},center}
        \begin{tikzpicture}
            \node[style = {rectangle, draw=blue, fill = blue!5, very thick, minimum width = 60mm, minimum height =20mm, align = center}] (ISM) {\Large ISM\\ \\ \\};
            
            \node[](white)[right=10mm of ISM]{};
            
            \node[style = {rectangle, draw=blue, very thick, minimum width = 15mm, minimum height =10mm)}] (cold)[below left=-11mm and -15.8mm of ISM]{cold};
            
            \node[style = {rectangle, draw=red, very thick, minimum width = 15mm, minimum height =10mm)}] (hot) [below right=-11mm and -16mm of ISM] {warm};

            \node[](coldlow)[below=-2mm of cold]{};
            \node[](hotlow)[below=-2mm of hot]{};
            
            \node[style = {rectangle, draw=blue, very thick, minimum width = 30mm, minimum height =30mm)}] (CGM) [left = 33mm of ISM] {\Large IGM/CGM};
            
            \node[](CGMlow)[below=-2.5mm of CGM]{};
            
            \node[style = {rectangle, fill = white, draw = black, align=center}](SF)[below = 15mm of cold]{star\\formation};
            
            \node[style = {rectangle, draw=yellow, fill = white, very thick, minimum width = 15mm, minimum height =10mm)}](Stars) [below  = 15 mm of SF]{Stars};
            
            \node[style = {rectangle, fill=white, draw=yellow, very thick, minimum width = 15mm, minimum height =10mm, align=center}](AGB) [below right  = 0 mm and 10mm of Stars]{AGB\\stars};

            \node[style = {rectangle, draw=yellow, very thick, minimum width = 15mm, minimum height =10mm)}](ccSN) [below left  = 0 mm and 10mm of Stars ]{ccSN};
            
            \node[style = {rectangle, draw=yellow, very thick, minimum width = 15mm, minimum height =10mm)}](NSM) [below left  = 0 mm and 10mm of ccSN]{NSM};

            \node[style = {rectangle, draw=yellow, very thick, minimum width = 15mm, minimum height =10mm)}](SNIa) [below right = 0 mm and 10mm of AGB]{SNIa};

            \draw [->] (cold) to[out=-15, in=195]
            node[ black, pos=0.5, above] {evaporation} (hot);
            
            \draw [very thick,<-] (cold) to[out=15, in=165]
            node[black, pos=.5, above] {condensation} (hot);

            \draw [-, very thick] (cold.south) -- (SF.north);
            \draw [->, very thick] (SF.south) -- (Stars.north);

            \draw [->] (Stars.south) --node[sloped, midway, text width=0.5cm]{\footnotesize light stars}(AGB.west);
            \draw [->] (Stars.south)--node[ sloped, anchor=0.1, text width=0.5cm]{\footnotesize massive$\;$ stars} (ccSN.east);
            \draw [dashed,->] (ccSN.south) -- node[sloped, anchor=center, below]{\footnotesize progenitor}(NSM.east);
            \draw [dashed,->] (AGB.south) -- node[sloped, anchor=center, below]{\footnotesize progenitor}(SNIa.west);

            \draw [->, very thick] (CGM.east) -- node[sloped, anchor=center,above]{\footnotesize inflow} (cold.west);

            \node[style = ellipse,fill =white, draw = green, very thick,align = center](AGBgas)[above =5mm of AGB]{s-process};
            
            \draw [->] (AGB.north) -- (AGBgas.south);
            \draw [->, shorten >= 1mm] (AGBgas.north) -- (coldlow.south);
            \draw [->, shorten >= 1mm] (AGBgas.north) -- (hotlow.south);

            \node[style = ellipse, fill = white,draw = green, very thick,align = center](ccSNgas)[above =5mm of ccSN]{$\alpha$-rich\\Mg};
            
            \begin{scope}[on background layer]
            \draw [->] (ccSN.north) -- (ccSNgas.south);
            \draw [->, shorten >= 1mm] (ccSNgas.north) -- (coldlow.south);
            \draw [->, shorten >= 2.2mm] (ccSNgas.north) -- (hotlow.south);
            \end{scope}
            
            \node[style = ellipse, fill = white,draw = green, very thick,align = center](NSMgas)[above =15mm of NSM]{r-process \\Eu};
            
            \begin{scope}[on background layer]
            \draw [->] (NSM.north) -- (NSMgas.south);
            \draw [->, shorten >= 1mm] (NSMgas.north) -- (coldlow.south);
            \draw [->, shorten >= 2.2mm] (NSMgas.north) -- (hotlow.south);
            \end{scope}
            
            \node[style = ellipse, draw = green, very thick,align = center](SNIagas)[above =15mm of SNIa]{Fe-rich};
            
            \begin{scope}[on background layer]
            \draw [->] (SNIa.north) -- (SNIagas.south);
            \draw [->, shorten >= 1mm] (SNIagas.north) -- (coldlow.south);
            \draw [->, shorten >= 1mm] (SNIagas.north) -- (hotlow.south);
            \end{scope}

            \begin{scope}[on background layer]
            \draw[dotted,thick,->, shorten >= 1.1mm] (ccSNgas.north) --  (CGMlow.south);
            \draw[dotted,thick,->, shorten >= 2mm] (AGBgas.north) --  (CGMlow.south);
            \draw[dotted,thick,->, shorten >= .7mm] (NSMgas.north) --  (CGMlow.south);
            \draw[dotted,thick,->, shorten >= 2mm] (SNIagas.north) --  (CGMlow.south);
            \end{scope}

        \end{tikzpicture}
    \end{adjustbox}
    
    \caption{Diagram showing the main gas flows and enrichment processes within a galaxy.}
    \label{fig:gasBalances}
\end{figure}

Tinsley's concept of abundance planes relies on the fact that different progenitor types like core collapse supernovae (ccSNe), SNIa, neutron star mergers, asymptotic giant branch stars (AGB-stars), etc., have both different typical timescales and different abundance patterns in their ejecta  \citep[see e.g.][]{rauscher_origin_2010, nomoto_nucleosynthesis_2013}. We illustrate the main processes in Fig.~\ref{fig:gasBalances} and give a summary of the assumed timescales in table \ref{tab:TimeScaleValues}.

\subsubsection{Elemental Production With Different Timescales}

Massive stars ending in ccSNe are the main producers of $\alpha$-elements like O, Ne, Mg, and contribute a major fraction of the Galactic iron peak elements around $^{56}$Fe \citep{hoyle_nuclear_1954, burbidge_synthesis_1957}. Before the supernova explosion, these massive stars ($\gtrsim 10 \,{\rm M}_\odot)$ develop an onion-like structure with more progressed and hence heavier burning regions in the centre. Depending on how much of their envelopes are stripped before the supernova \citep[up to being Wolf-Rayet-stars,][]{crowther_physical_2007}, they are classified as SNII (hydrogen lines)  and SNIb,c (no hydrogen lines, with/without helium), collectively called core collapse supernovae\footnote{We do not differentiate exotic types here like symbiotic massive systems, electron capture supernovae or hypernovae}. The high mass drives fast nucleosynthesis, exhausting the stellar fuel at of order a few Myrs and making enrichment with $\alpha$-elements like magnesium (Mg) the fastest enrichment process \citep[e.g.][]{tinsley_stellar_1979, bressan_evolutionary_1993}.  

Massive stars in multiple systems can leave behind tight neutron star binaries after undergoing a ccSN. When these merge, the ejected matter decompresses and undergoes r-process nucleosynthesis. In this process, neutrons are rapidly captured onto a seed nucleus under extremely high neutron fluxes until it travels along the neutron drip line. After the capture phase, it $\beta$-decays back to stability.  Higher densities where the magic neutron numbers cross the drip-line result in the typical r-process abundance peaks. This produces elements like Au, U, Th, Gd, Eu, etc. \citep[see e.g.][for a recent review]{ cowan_origin_2021}.
The site of the r-process has been debated. 
After the first detected neutron star merger  \citep[NSM;][]{abbott_gravitational_2017}, r-process material was found at the explosion site \citep{kasen_origin_2017, watson_identification_2019, chornock_electromagnetic_2017,rosswog_first_2018}, proving that r-process does take place in NSMs. 
However, other contributing pathways have been proposed, especially  as observed [$\alpha$/Fe] abundances in the Milky Way did not fit with the expected timescales \citep{cote_origin_2018, cote_neutron_2019}.
Following this, \citet{simonetti_new_2019} suggested that this can be rectified when assuming a metallicity dependent NS binary occurrence rate. Using a non-chemistry based approach, \citet{beniamini_gravitational_2019} studied 15 gravitational wave detections of NSMs and concluded that at least about half of the NSMs occur without significant time delay after star formation. In contrast, \citet{molero_origin_2023} concluded that a significant r-process contribution of SNe is necessary to fit the solar neighbourhood observations. 
However, neither of those take into account a two phase interstellar medium, i.e. a hot and a cold gas phase.
 \citet{schonrich_chemical_2019} showed that using a two-phase ISM (see Sec.~\ref{sec:multiphasegas}) immediately brings the models into agreement with the observations.
 Additionally, collapsars, a major contender for the dominant r-process site \citep{siegel_collapsars_2019} was ruled out in \citet{fraser_metallicity-suppressed_2021}.
Hence, for this work we consider NSMs as the sole r-process site.\footnote{We also tested a delay time distribution consistent with \citet{beniamini_gravitational_2019}. However, except for a small scaling effect on the europium abundance it has no effect on our results.} 
As neutron star binaries require time for the inspiral (a too small initial separation will likely be disfavoured due to the problem of inspiral in common envelope evolution), the NSM time delay distribution resembles the SNIa time delay distribution (see below) but with timescales that are about an order of magnitude shorter.
Here we use a delayed onset time of 20 Myr and a decay distribution of ${\rm e}^{-t/\rm 300 Myr}$.\footnote{ Our choice is at the upper end of the accepted range \citep{hotokezaka_neutron_2018}, but the parameter is not substantative to our qualitative discussion, as it just scales the temporal/spatial separation between ccSN and NSM, i.e. the position of peak Eu production. } 

At lower neutron fluxes than the r-process, the s-process contributes most elements between iron and lead. It likely does take place in (rotating) massive stars \citep{pignatari_s-process_2008,frischknecht_non-standard_2012}, but the majority is found to stem from intermediate stars in their AGB-phase \citep{clayton_neutron_1961}. Particularly, elements with "magic" neutron numbers in the valley of stability are produced by the s-process (e.g. Pb). Those have a corresponding peak of r-process elements (near Au for Pb) at lower nucleon numbers, due to the different neutron to proton ratio at which r- and s-process yield a magic neutron number isotope \citep{sorlin_nuclear_2008}. 
While we do not explicitly track s-process elements in this paper, our evolution model does fully incorporate yields from low to intermediate mass stars. Beyond their s-process contribution, they contribute helium enrichment and larger amounts of otherwise unenriched yields at higher population ages and hence are an important source of low-metallicity gas.

Finally, the majority of iron peak elements are produced by white dwarfs (WDs) exploding as SNIa \citep{thielemann_explosive_1986}. There are many possible channels for SNIa: classic single degenerate scenarios where an accreting CO WD exceeds the Chandrasekhar mass at $\sim 1.38\,{\rm M}_\odot$ \citep[][]{chandrasekhar_highly_1931, whelan_binaries_1973,nomoto_accreting_1984}, mergers or collisions of two WDs \citep[the double degenerate scenario, e.g. ][]{iben_supernovae_1984, iben_evolutionary_1987}, induced explosion by a helium nova on the surface \citep{fink_double-detonation_2010,kromer_double-detonation_2010,jiang_hybrid_2017}, and others. Likely, all channels contribute \citep[see e.g.][]{wang_progenitors_2012, livio_progenitors_2018, soker_supernovae_2018}. We can afford to be agnostic on this issue, as chemical evolution relies mainly on the feedback time. Here we assume standard yields from \citet{iwamoto_nucleosynthesis_1999}. After an initial delay time of 0.45$\,{\rm Gyr}$ we explode the SNIa on two exponential delay distributions, where a very small fraction of currently 1$\%$ follows a short timescale with ${\rm e}^{-t/\rm 100 Myr}$ and  99 $\%$ a longer decay time of ${\rm e}^{-t/\rm 1.5 Gyr}$. In order to not add another free parameter, we do not consider a prompt fraction of early exploding SNIa as proposed by \citet{palicio_analytic_2023}.
We also note that the progenitor stellar remnants, i.e the white dwarfs from ccSNe and neutron stars from AGB-stars, are already included in our scattering framework before their final nucleosynthesis stage as SNIa and NSM. This ensures that those yields are significantly more dispersed than the ccSN yields that closely follow the radius of their initial formation.

\subsubsection{Different gas phases}
\label{sec:multiphasegas}
Another enrichment delay comes from the two-phase nature of the ISM. 
ISM models often distinguish three or more phases \cite[see e.g.][]{ferriere_interstellar_2001}. Contrasting with this most chemical enrichment models make the simplifying assumption of one gas phase. However, it is well-established that galaxies are surrounded by a hot circumgalactic medium (CGM), that most SN yields are initially in the hot phase (see e.g. the crab nebula), and that the hot ISM is cooling down slowly to feed the cold star-forming ISM on a timescale of ${\sim} 1 \,{\rm Gyr}$. Additionally, observations show both warm winds and chimney-like structures emanating from discs by which the hot-phase yields from SN escape from the disc plane \citep[e.g.][]{krause_galactic_2021}. Hence, incorporating at least the effects on elemental abundances of a multiphase medium is an important part of a chemical evolution model.

 One reason the cooling delay \citep[first implemented in classical chemical evolution by][]{schonrich_chemical_2009} has been neglected in chemical evolution models is that, as demonstrated by \citet{spitoni_effects_2009}, its impact becomes significant only when metal-dependent yields are considered. Once these are taken into account however, it becomes clear that a two-phase ISM is crucial in chemical evolution. For instance, \citet{schonrich_understanding_2017} showed that time delay plays an essential role in reversing disc metallicity gradients \citep[e.g.][]{spagna_evidence_2010}.

Most importantly, the delay from the hot ISM directly resolves the r-process enrichment problem, as different splitting of yields into the cold vs. hot gas phases can effectively invert the time-delay relationship between r-process elements from NSM and $\alpha$-elements from ccSN \citep{schonrich_chemical_2019}. This study resolved the stark conflicts between chemical evolution models and observational evidence for neutron star mergers \citep{cote_origin_2018,cote_neutron_2019, chornock_electromagnetic_2017}, if the fraction of NSM yields injected into the cold phase is a factor ${\sim} 2$ larger than the fraction of ccSN yields. This argument was further sharpened by \cite{fraser_metallicity-suppressed_2021}. While there are good reasons to expect different fractions (NSM are spatially and temporally less correlated with the energetic outflows from ccSN that eject material far from the star-forming regions, and the heavy elements they yield cannot commonly be ionised in the hot ISM, which leads to faster line cooling), the exact splitting into different phases lacks independent measurement paths, especially in the outer Galactic disc with its fairly continuous star formation.

\subsection{Nuclear Stellar Discs}
\subsubsection{General Structure and Formation}
\label{sec:nucleardiscs}

Like the majority of disc galaxies, the Milky Way possesses a central Galactic bar dominating the inner $\sim 4 \,{\rm kpc}$. It was first inferred from gas motions in radio observations \citep[][]{johnson_is_1957}. Still, due to the heavy obscuration of the Galactic central regions, it has only been widely accepted about 30 years ago, following \citet{blitz_direct_1991} and \citet{binney_understanding_1991}, through extended studies on radial gas inflow \citep[e.g.][]{fux_3d_1999, englmaier_gas_1999} and closer examination of the Galactic bulge that extends over the disc \citep{stanek_color-magnitude_1994, binney_photometric_1997}. More recently, the stellar bar was clearly identified by its boxy/peanut shape \citep{mcwilliam_two_2010, nataf_split_2010}.

Bars naturally form in most galaxies. Especially in galactic centres, even between the resonances, stellar orbits are strongly shaped by the bar potential and their orientation with respect to the long axis of the bar changes. It can be shown \citep[see e.g. chap. 3.3 of][]{binney_galactic_2008} that inside the inner Lindblad resonance (ILR), the stellar orbits (so-called x$_2$-orbits) are anti-aligned with the bar, while the bar itself is formed (amongst other orbit families) by the elongated x$_1$-orbits between the ILR and the co-rotation resonance. This also poses a constraint to the possible length of the bar, as the orbits further out are again anti-aligned with the bar. Indeed, there are some reports about an inner ring in the Milky Way hypothesized to be formed by those orbits just outside co-rotation \citep{wylie_milky_2022}.

As the bar sweeps through the gas in the galactic disc, it creates long shock lanes along its leading edges and tips \citep[see e.g. fig. 3 of][]{athanassoula_existence_1992}.  This gas falls on the leading side of the bar into the central region and accumulates on the dense NSD \citep[on x$_2$ orbits,][]{contopoulos_orbits_1989}.

Turbulence and shear suppress star formation within the bar lines \citep[e.g.][]{kim_impacts_2024, maeda_galactic_2025}, however, recent observations showed this effect to not be as strong as initially expected \citep{diaz-garcia_molecular_2021, querejeta_stellar_2021, maeda_statistical_2023}. Nevertheless, as the inflow speed is of order $\gtrsim 100 \,{\rm km}\,{\rm s}^{-1}$ (as found e.g. by \citet{sormani_fuelling_2023} for NGC 1097), the residence time within the bar is quite short ($\sim $ a few 10 Myr) and should hence not strongly influence the NSD chemistry. We hence model the bar lane accretion as direct onfall from the bar tips.

Predicting the size of an NSD is difficult. Historically, it was only known that the ILR poses an outer limit \citep[see e.g.][]{simkin_nearby_1980, combes_spiral_1985, buta_galactic_1996}. Advances have been made here, taking e.g. the viscosity/sound speed of the gas into account \citep[e.g.][]{van_de_ven_migration_2009, sormani_dynamical_2018}. 
Recently, \citet{sormani_nuclear_2024} proposed that nuclear rings are formed by gas accumulating at the inner edge of a gap near the Inner Lindblad Resonance (ILR) of a bar potential. This gap is created and then widened by trailing waves, which are continuously excited by the bar and remove angular momentum from the nuclear gas disc.

The radial region at which the inflowing gas settles in the galactic centre is referred to as nuclear ring (for the MW this is the central molecular zone (CMZ)). Due to the strongly increased gas density in this region, the nuclear ring vigorously forms stars  \citep[up to $5 \%$ of its galaxy's star formation;][]{kennicutt_demographics_2005}. These rings were observed \citep{knapen_circumnuclear_1999,comeron_ainur_2010} before being shown to be the outer edge of inside-out forming, i.e. radially growing NSDs \citep{bittner_inside-out_2020}. This was corroborated by the detection of a consistent radial age gradient \citep{nogueras-lara_evidence_2023}. We thus concentrate on inside-out forming NSD models in this paper.

A consistent gas supply to the NSD is ensured by the deceleration of the galactic bar, due to loss of angular momentum to the dark halo \citep[see e.g.][]{debattista_constraints_2000,berentzen_gas_2007,dubinski_anatomy_2009}. This slow-down moves the bar resonances outwards and hence also the radius from which the bar funnels material to the centre. In case of the MW, this deceleration has recently been proven in solar neighbourhood kinematics \citep{friske_more_2019,chiba_resonance_2020,chiba_tree-ring_2021}, which also allowed for a precise measurement of the MW bar pattern speed and corotation resonance radius ($R_{CR} = 6.6 \pm 0.2 \,{\rm kpc}$). The outward shift in the feeding radius \citep[expected to be roughly linear in time from simulations; see e.g. Fig.2 in ][]{aumer_origin_2015} strongly affects the composition of funnelled-in central gas. We will discuss this in Sec.~\ref{sec:Results}.

Due to their extremely high density and unique formation process, NSDs are fascinating objects of study, with the MW NSD being the closest and hence most accessible example. However, its central position also makes direct observations difficult. Due to the heavy dust obscuration ($\gtrsim 3 \,{\rm mag}$ in the infrared K-band, completely blacking out the V-band) and  measuring only $\approx 150 \,{\rm pc}$, the identification of the CMZ \citep[as slowly pieced together as it was][]{rougoor_distribution_1960, petters_models_1975, scoville_survey_1972, bania_carbon_1977,liszt_gas_1978,bally_galactic_1988} predated the stellar observation in first density and then kinematics \citep{launhardt_nuclear_2002, schonrich_kinematic_2015} by nearly half a century. To date, detailed and reliable abundance information for the MW NSD is scarce (see also App.~\ref{sec:dataSituation}). We also note that the relationship between the MW nuclear star cluster (NSC, ${\sim} 100$ times smaller) and the NSD, in particular the contribution of later-formed nuclear disc stars to a pre-dating nuclear star cluster is unknown and we will leave discussion to future study. 

Finally, it is important to discern NSDs from other strongly light-emitting objects at the centre of galaxies, especially from the inherently older classical bulges or X-shaped bars, which are either caused by (recurrent) bar buckling \citep{athanassoula_morphology_2002, martinezvalpuesta_evolution_2006, fragkoudi_ridges_2019}, or secular processes \citep{sellwood_three_2020}. In contrast to this, nuclear disc stars form after the creation of the bar \citep[see][for a discussion]{baba_age_2020} in situ in the NSD. Compared to classical bulges, NSDs will have much stronger rotational support and likely colder kinematics.

We note that a small number of NSDs have been detected in non-barred galaxies \citep[see e.g. the AINUR catalogue][]{comeron_ainur_2010}. However, these are rare, comparably faint, and might result from merger events \citep{MayerFormationMergers} and/or destruction of their host bars. We leave these to a a later study.

\subsubsection{Chemical Composition}
\begin{figure}
    \centering
    \includegraphics[width =\columnwidth]{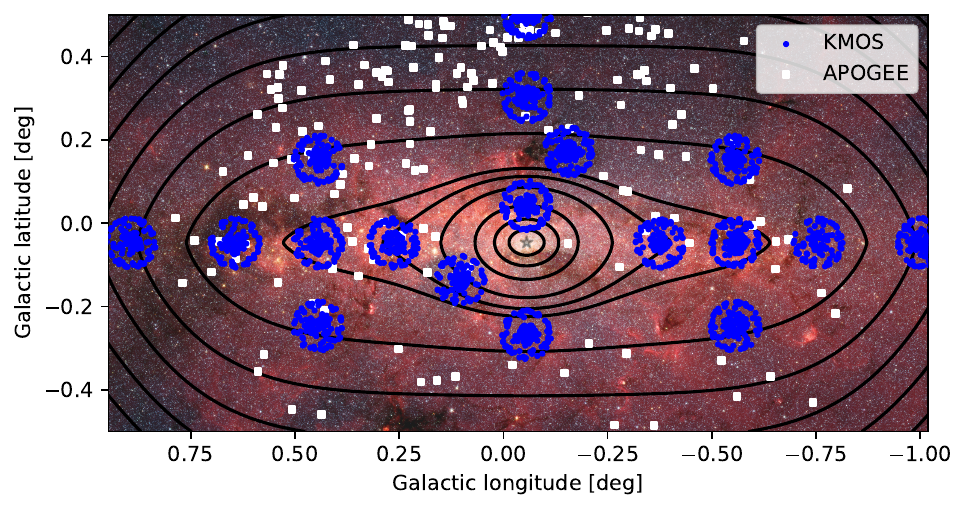}
    \caption{View of the Galactic Centre of the Milky Way. Background: False colour Spitzer image (NASA/JPL-Caltech/S. Stolovy (Spitzer Science Center/Caltech)). Overlayed are APOGEE DR17 \citep{abdurrouf_seventeenth_2022} NSD star passing recommended quality checks (see also Sec.~\ref{sec:appendixExtraMaterial}) with AK$\_$WISE $> 2$ as well as the KMOS fields from \citet{fritz_kmos_2021}. The contours show the composite best fitting stellar densities of NSD from \citet{launhardt_nuclear_2002} and nuclear stellar cluster from \citet{chatzopoulos_old_2015}.}
    \label{fig:GalacticCentre}
\end{figure}

As mentioned above, our in-disc position hampers observations of the MW NSD. It is hence beneficial to also study other galaxies, that, with high-enough inclination provide a spatially resolved view. Currently, observed NSDs tend to be larger ($500$ pc vs. $\sim 150$ pc for the MW) and more dominant in mass and luminosity. 

The Calar Alto Legacy Integral Field Area Survey (CALIFA) has spectroscopically mapped $\sim 600$ local galaxies. \citet{sanchez-blazquez_stellar_2014} studied the bar influence in $62$ face-on spiral galaxies. Some of these show an NSD; however, the resolution is not sufficient to spatially dissect the NSDs and quantitatively compare with the bar tips. CALIFA has also only studied metallicity and not detailed abundance ratios.

The BaLROG (Bars in Low Redshift Optical Galaxies) sample taken with the Spectrographic Areal Unit for Research on Optical Nebulae  \citep[SAURON, William Herschel Telescope, Observatorio del Roque de los Muchachos, La Palma;][]{bacon_sauron_2001} comprises $16$ large mosaics of barred galaxies. \citet{seidel_balrog_2016} calculate ages, metallicities and [Mg/Fe] abundance ratios mapping their gradients along the bar major and minor axes. Again, the spatial resolution precludes definitive conclusions; however, visual inspection comparing  the NSD abundances to the abundance at the bar tips at least suggests a correlation. 

Explicitly dedicated to NSDs and nuclear rings is the TIMER (Time Inference with MUSE in Extragalactic Rings) sample of $21$ barred spiral galaxies taken with the MUSE (Multi Unit Spectroscopic Explorer) integral field spectrograph \citep{Gadotti2020NucDiskBar}. \citet{bittner_inside-out_2020} corroborated the idea of an analogous inside-out formation of NSDs to ordinary galactic discs and of nuclear rings as the most recent place of gas accretion and star formation at the rim of their NSD. They also observe enhanced $\alpha$-enrichment in nuclear rings. However, their observations do not reach out to the ends of the host bars, preventing a direct determination of the nucleosynthesis within the NSD in comparison to the material falling in from the bar tips.

Also using MUSE, the PHANGS (Physics at High Angular resolution in Nearby GalaxieS)-MUSE survey \citep{emsellem_phangs-muse_2022} studies $19$ galaxies, most of which harbour NSDs. The first data release did not yet provide reliable metallicities. (A first study, deriving metallicities \citep{groves_phangs-muse_2023} showed  global metallicity variations, but from H$_\alpha$ emitting, ionised nebulae that are not well-resolved within the NSD.) When future data releases include detailed abundances, we expect these data the best route to constrain our models.

Looking back at our own Galaxy, the Apache Point Observatory Galactic Evolution Experiment \citep[APOGEE, see][]{majewski_apache_2017} has a few pointings in the Galactic centre. The latest and final data release DR17 \citep{abdurrouf_seventeenth_2022} contains spectra for 43,200 bulge stars. A unique NSD selection is difficult, but using a reddening cut of $A_K > 2.0 \,{\rm mag}$ yields ${\sim} 300$ very bright, but cool ($T< 3600$ K) giant stars. However, while the spectra look promising for a dedicated analysis, using the prederived stellar parameters, only 113 stars with published metallicities pass the recommended quality cuts (see App.~\ref{sec:appendixExtraMaterial} for details). Thus selected stars show generally  low metallicities (down to $\hbox{[Fe/H]}\xspace{\sim}-1.2$ and one outlier at $\hbox{[Fe/H]}\xspace{\sim}-2.5$), which are significantly lower and also strangely distributed compared to e.g. the KMOS NSD sample discussed below. Interestingly, especially the stars with $A_K > 3.0 \,{\rm mag}$,  used in \citep[][]{schonrich_kinematic_2015} for the detection of the NSD due to their strong rotational signature, seem to have particularly low metallicities.  
(See also Fig.~\ref{fig:ApogeeKmos}.)

Recently, major advances are made using the K-band Multi Object Spectrograph \citep[KMOS;][]{sharples_first_2013} operating on the VLT that is specifically designed for integral field spectroscopy at high reddening. \citet{fritz_kmos_2021} showed for the first time spectroscopic metallicities for $\sim 3000$ NSD stars, obtained by \citet{feldmeier-krause_kmos_2017}, again mostly targeting RGB-stars, so age and metallicity selection biases will have to be taken into account. \citet{schultheis_nuclear_2021} used this data for a first study of metallicity gradients in the Galactic NSD. They found that the NSD is likely made up by a larger, kinematically cold, rotating metal rich population and a smaller, partially even retrograde metal poor sample. (Likely the latter is made up of of bar orbits that can appear retrograde if projected onto the smaller NSD.) They measured a vertical gradient, which they suspect to be due to vertically increasing contamination with bulge stars, and found no significant radial metallicity gradients. However, the strong favouring of nuclear ring stars by sample selection and dust obscuration is not yet accounted for. 
Very recently, \citet{nogueras-lara_smooth_2023} connected those to photometric data and indeed found a negative metallicity gradient consistent with extragalactic results e.g. in \citet{bittner_inside-out_2020}.
The pointings of APOGEE and the KMOS NSD survey in the MW NSD can be found in Fig.~\ref{fig:GalacticCentre}.

\section{Model}
    \label{sec:Model}
     We use the RAMICES II GCE model described in \citet{fraser-govil_advancements_2022} and also used in \citet{fraser_metallicity-suppressed_2021}.  
    It follows the same concepts as the chemical evolution models from \citet{schonrich_chemical_2009}, with added inside-out formation \citep{schonrich_understanding_2017} and the hot/cold gas phase calibrations from \citet{schonrich_chemical_2019}. The code will be published separately, but can already be retrieved.\footnote{\url{https://github.com/DrFraserGovil/RAMICES_II}} In Sec.~\ref{sec:GeneralModel} we first outline the main elements of the model, before detailing the changes necessary for including an NSD in our new adaptation in Sec.~\ref{sec:ModellingBarAndNSD}. The main idea is to run two coupled RAMICES II models, one for a full galactic disc whose abundances then inform the accretion for the second model, resolving the NSD. This ensures a single, self‐consistent set of chemical evolution parameters throughout the simulation while affording higher spatial resolution in the NSD than is typical for full‐galaxy simulations.
    \subsection{Galactic Model}
    \label{sec:GeneralModel}
      	
      	\begin{table}
    	    \centering
    	    \caption{Main Simulation parameters for the galactic simulation. The definition for the infall is the same as in Equation 2 of \citet{schonrich_chemical_2009}.}
    	    \begin{tabular}{cc}
                \hline\hline
            \textbf{Parameter} &\textbf{ Value} \\
            \hline
    	         simulation time/ age &   12 Gyr\\
    	         step size & 10 Myr \\
    	         galactic disc ring number & 100 \\
    	         galactic disc ringwidth & 0.2 kpc\\
    	         total mass at start ($M_0$)& $10 ^{8} \,{\rm M}_\odot$\\
    	       final scalelength ($R_f$) & 4 kpc \\
                  early infall mass ($M_1$) & $40\times 10 ^9 \,{\rm M}_\odot$\\
                  early infall timescale ($b_1$)& $1 \,{\rm Gyr}$\\
                  long infall mass ($M_2$)& $100 \times 10 ^9 \,{\rm M}_\odot$\\
                  long infall timescale ($b_2$)& $9 \,{\rm Gyr}$\\
            \hline\hline
    	    \end{tabular}
    	    \label{tab:simparam}
    	\end{table}

    As the currently best NSD abundance measurements stem from the Milky Way, the models in this work will be tailored to MW like galaxies and NSDs. 
    Table \ref{tab:simparam} lists the main set-up parameters for the galactic model used throughout this paper, while a further description of used timescales and yields can be found in Sec.~\ref{sec:appChemEv}.
    
    The simulation uses (time-dependent) disc scalelengths (see eq.~\eqref{eq:scalelength} and Fig.~\ref{fig:scalelength}), total galactic stellar mass (${\sim} 5 \times 10^{10} \,{\rm M}_\odot$), and total age ($12 \,{\rm Gyr}$) comparable to the Milky Way. With timesteps of $10 \,{\rm Myr}$, it resolves 100 concentric rings (width $200 \,{\rm pc}$), which yields sufficiently small abundance differences between neighbouring rings ($\sim 0.01 \,{\rm dex}$). 
    
    We use a Chabrier initial mass function \citep[IMF, ][]{chabrier_galactic_2003}. With the difference to e.g. Salpeter's IMF mostly on low-mass stars, the IMF choice mainly results in moderate equilibrium metallicity differences, which are degenerate with a mass loss parameter. Star formation rates (SFR) are determined with a Kennicutt-Schmidt law \citep{kennicutt_jr_global_1998, schmidt_rate_1959}: Star formation efficiency rises at higher gas surface density, with a cut-off at low density where the gas becomes stable against gravitational collapse. This is important especially in a region like the NSD with high gas density variations.
    We note that it is likely that magnetic fields \citep[e.g.][]{moon_effects_2023} and turbulence \citep{nesvadba_dense_2011,longmore_variations_2013} change the SFR in the NSD and specifically in the CMZ.

\subsection{ISM, Gas Flows}    

    \begin{table}
        \centering
        \caption{Fiducial NSD and chemical evolution Model parameters}
        \begin{tabular}{cc}
        \hline\hline
        \textbf{Parameter} & \textbf{Values}\\
        \hline
             NSD ring number & 50\\
             NSD ringwidth & 3 pc \\
           total mass at start & $10 ^{6} \,{\rm M}_\odot$\\
           final scalelength & 70 pc \\
           \hline
             $ f_{\mathrm{c,ccSN}}$ &   0.2\\
             $ f_{\mathrm{c,NSM}}$ & 0.4 \\
             $ f_{\mathrm{c,AGB}}$ & 0.3 \\
             $ f_{\mathrm{c,SNIa}}$ & 0.01\\
             global eject fraction & 0.55\\
             nuclear eject fraction & 0.65 \\
             
        \hline \hline

        \end{tabular}
        \label{tab:simparam_bestfit}
    \end{table}

    \begin{figure}
        \centering
        \includegraphics[width = 0.9\columnwidth]{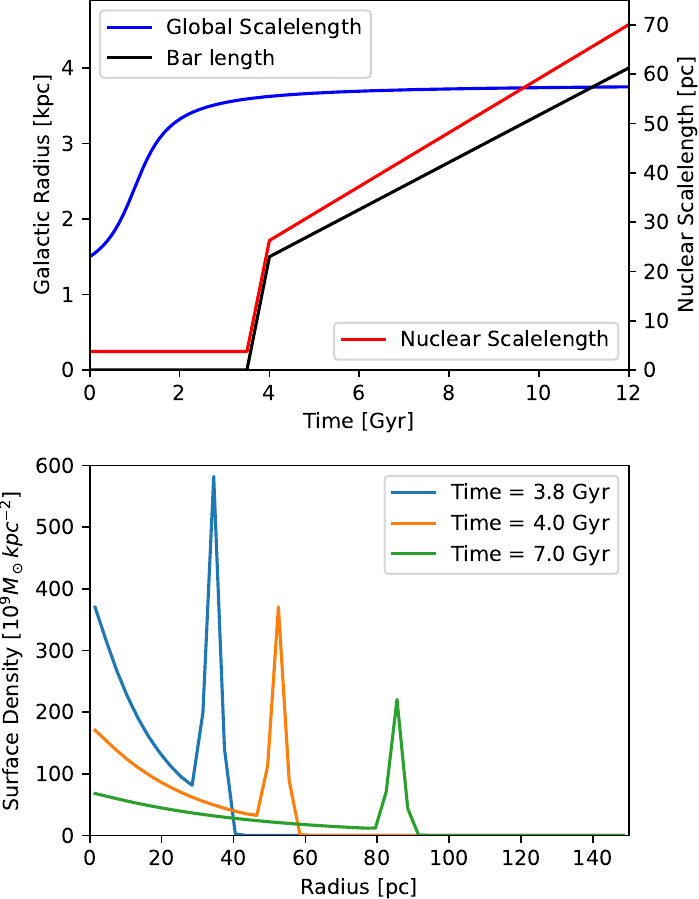}
        \caption{\textit{Top:} Evolution of the whole galactic gaseous disc (blue) and nuclear gas disc (red) scale length and the assumed bar length (black) over simulated time.\\
        \textit{Bottom:} Shape of the prescribed gas distribution in the NSD at different times.}
        \label{fig:scalelength}
    \end{figure}

    The multi-phase ISM is implemented as a hot and a  cold (star forming) reservoir. The different contribution of each feedback source $X$ to the cold gas phase, $f_\text{c,X}$, are listed in table \ref{tab:simparam_bestfit}. They are chosen consistently with \citet{schonrich_chemical_2019}, such that $\ensuremath{ f_{\mathrm{c,NSM}}}/\ensuremath{ f_{\mathrm{c,ccSN}}} \sim 2$ and the remaining freedom is used to improve the match to the local abundance gradients. However, those proposed values should only be seen as a starting point for future analysis. 
    
    The initial galactic gas disc only contains primordial, cold gas with total mass $10^8 \,{\rm M}_\odot$ as $76\%$ hydrogen and $24\%$ helium. The galactic disc forms inside-out with a growing gas scalelength (Fig.~\ref{fig:scalelength}) that settles at $3.77 \,{\rm kpc}$ at the upper end of current observations \citep[][] {kalberla_global_2008}. 
    
    Following \citet{schonrich_chemical_2009} we parametrise the gas onfall onto the galaxy with two exponential functions in time, one fast component driving the initial build-up of the disc and the slower term guiding the long-term decline in star formation. The parameters match \citet{schonrich_understanding_2017} apart from a higher fast onfall mass of $40 \times 10^9 \,{\rm M}_\odot$ (see Tab. \ref{tab:simparam}). RAMICES II approximates angular momentum conserving gas accretion via the hot galactic corona \citep[see e.g.][]{marinacci_mode_2010}: due to the corona's thermal pressure support, the onfalling gas has a lower specific angular momentum (between  ${\sim} (2/3)$ and $(3/4)$ of the circular velocity $v_{\mathrm{c}}$) than the disc gas (${\sim} v_{\mathrm{c}}$); thus accretion dilutes disc angular momentum, driving a continuous radial inflow. 
   While it is unclear how important this effect is for large scale galactic discs, it will likely be much more relevant for the small, turbulent and quite magnetised NSDs. 
    \citet{portinari_radial_2000} argued that a galactic bar not only shapes the chemical evolution in the area it sweeps over, but impacts the global metallicity gradient and gas profile as it quickly takes up gas around its tips and efficiently funnels it towards the centre.
    Lastly, large scale structures such as spirals and bars drive angular momentum exchange and inflow. In total, the resulting flows determine the global metallicity gradient \citep{lacey_chemical_1985,goetz_abundance_1992}.
    Here, we model gas inflow and onflow with the recipe of \citet{bilitewski_radial_2012}, which couples radial inflow, onfall and their radius-dependent ratios by balancing their angular momentum.

  Some gas will leave the galactic disc region: we assume that in each supernova, NSM, etc., a fraction of the yields is directly ejected to the CGM (we call it galactic disc ejection fraction in this paper). How this ejection fraction affects overall metallicity and chemical evolution timescales is tested in Sec.~\ref{sec:Results}.

\subsection{Modelling Bar and NSD}
\label{sec:ModellingBarAndNSD}

\subsubsection{Density Profiles and Accretion}

The simulation uses concentric rings and is hence axisymmetric, but the main effect of a bar on chemical evolution can be modelled by the radial gas flows it evokes, namely, wiping out the gas disc in the bar region, funnelling gas from the bar tips into the central region, and building up an NSD with high density and rapid star formation. 

Here we model this new NSD in a second, separate, simulation with 50 rings of width $3 \,{\rm pc}$, i.e. a radial extent of $150 \,{\rm pc}$. The NSD starts with negligible mass; once the bar forms, the NSD quickly receives the cold gas present within the bar radius.
After this, the bar length grows linearly and the galactic disc gas that is reached by this boundary is added onto the NSD.

To determine the distribution of onflow onto the NSD, we fix the shape of its gas surface density. We prescribe a two-part structure with equal mass in the exponential disc (with scalelength $R_{\mathrm{ND}}(t)$) and the nuclear ring at $r_{\mathrm{NR}}(t) = 2 R_{\mathrm{ND}}(t)$ (i.e. with a nuclear ring fraction $f_{\mathrm{NR}} = 0.5$). The nuclear ring has a Gaussian width $\sigma_{NR} = 0.35 \times 5 \,{\rm pc}$, and is followed at $r > R_{\mathrm{NR}}(t) + 5 \,{\rm pc}$ by a steep cut-off (communicated by $\delta_{\mathrm{cut-off}}$) with an exponential scale of $1 \,{\rm pc}$. 
This leads to a total surface density of
\begin{align}
    \label{eq:surface_density}
    \Sigma(r,t) &=    \left(1 - f_{\mathrm{NR}}\right)  \Sigma_{\mathrm{Disc}}  \delta_{\mathrm{cut-off}}(r, t)   + f_{\mathrm{NR}}  \Sigma_{\mathrm{NR}} \\ 
    \Sigma_{\mathrm{disc}}(r,t)& = \Sigma_0 \exp\left(-\frac{r}{R_{\mathrm{ND}}(t)}\right) \notag \\
    \Sigma_{\mathrm{NR}}(r,t) &= \frac {\Sigma_0}{\sigma_{\mathrm{NR}} {\sqrt {2\pi }}}\exp \left(-{\frac {(r-r_{\mathrm{NR}})^{2}}{2\sigma_{\mathrm{NR}} ^{2}}}\right) \notag \\
    \delta_{\mathrm{cut-off}}(r,t) &= 
    \begin{cases}
    1 & r< r_{\mathrm{NM}} + \Delta_{\mathrm{NR}}/2\\
    \exp\left(- \frac{r - (r_{\mathrm{NM}} + \Delta_{\mathrm{NR}}/2 )}{1 \,{\rm pc}} \right) & r \geq r_{\mathrm{NM}} + \Delta_{\mathrm{NR}}/2 \notag\\
    \end{cases},
\end{align}

where the scalelength as a function in time is
\begin{equation}
\label{eq:scalelength}
    R_{\mathrm{ND}}(t) = 
    \begin{cases}
    R_{{\mathrm{ND}}, 0} & t < t_{\mathrm{bar}}\\
    R_{{\mathrm{ND}}, 0} + \frac{t - t_{\mathrm{bar}}}{t_{\mathrm{i}}} (R_{\mathrm{ND, i}} - R_{{\mathrm{ND}}, 0}) & t_{\mathrm{bar}}{<} t {< }t_{\mathrm{bar}} {+} t_{\mathrm{ini}} \\
    R_{\mathrm{ND, i}} + \frac{t - (t_{\mathrm{bar}} + t_{\mathrm{i}}) }{t_{\mathrm{total}} - (t_{\mathrm{bar}} + t_{\mathrm{i}} )} (R_{\mathrm{ND, f}} - R_{\mathrm{ND, i}}) & t> t_{\mathrm{bar}} + t_{\mathrm{i}} $.$\\
    \end{cases}
\end{equation}
Here $R_{{\mathrm{ND}}, 0}$ is the starting gas scalelength at bar formation ($t_{\mathrm{bar}} = 3.5 \,{\rm Gyr}$), $R_{\mathrm{ND, i}}$ the gas scalelength of the NSD at the end of the initial growth phase, and $R_{\mathrm{ND, f}}$ the final gas scalelength (see Fig.~\ref{fig:scalelength}, which shows the scale- and barlength over time as well as the prescribed surface density).
The bar length follows the analogous equations:
\begin{equation}
\label{eq:barlength}
    r_{\mathrm{bar}}(t) = 
    \begin{cases}
    0 & t < t_{\mathrm{bar}}\\
    \frac{t - t_{\mathrm{bar}}}{t_{\mathrm{i}}} r_{\mathrm{bar,i}} & t_{\mathrm{bar}}< t < t_{\mathrm{bar}} + t_{\mathrm{i}} \\
    r_{\mathrm{bar,i}} + \frac{t - (t_{\mathrm{bar}} + t_{\mathrm{i}}) }{t_{\mathrm{total}} - (t_{\mathrm{bar}} + t_{\mathrm{i}} )} (r_{\mathrm{bar, f}} - r_{\mathrm{bar, i}}) & t> t_{\mathrm{bar}} + t_{\mathrm{i}} \\
    \end{cases}
\end{equation}
where $r_{\mathrm{bar, i}} = 1.5 \,{\rm kpc}$ is the length of the bar after the initial growth phase and $r_{\mathrm{bar, f}} = 4\,{\rm kpc}$ the final bar length.
With these definitions, the inner part of the NSD has a comparably low gas density at the end, dwarfed by the nuclear ring gas density. This is consistent with findings that the majority of the NSD extinction happens at the nuclear ring at the outer edge, while the NSD itself accounts only for about $10 \%$ of the total extinction \citep{nogueras-lara_first_2022}.

We make the assumption that the NSD inflow inherits the abundance at the bar tips. For this, cold gas from the radius of the extending bar tips is put into a reservoir (which may be thought of as the infall lanes) and immediately transferred onto the NSD conserving the density profile above.

\subsubsection{Transport Within the NSD}

   \begin{figure*}
        \centering
\includegraphics[width = 0.95\textwidth]{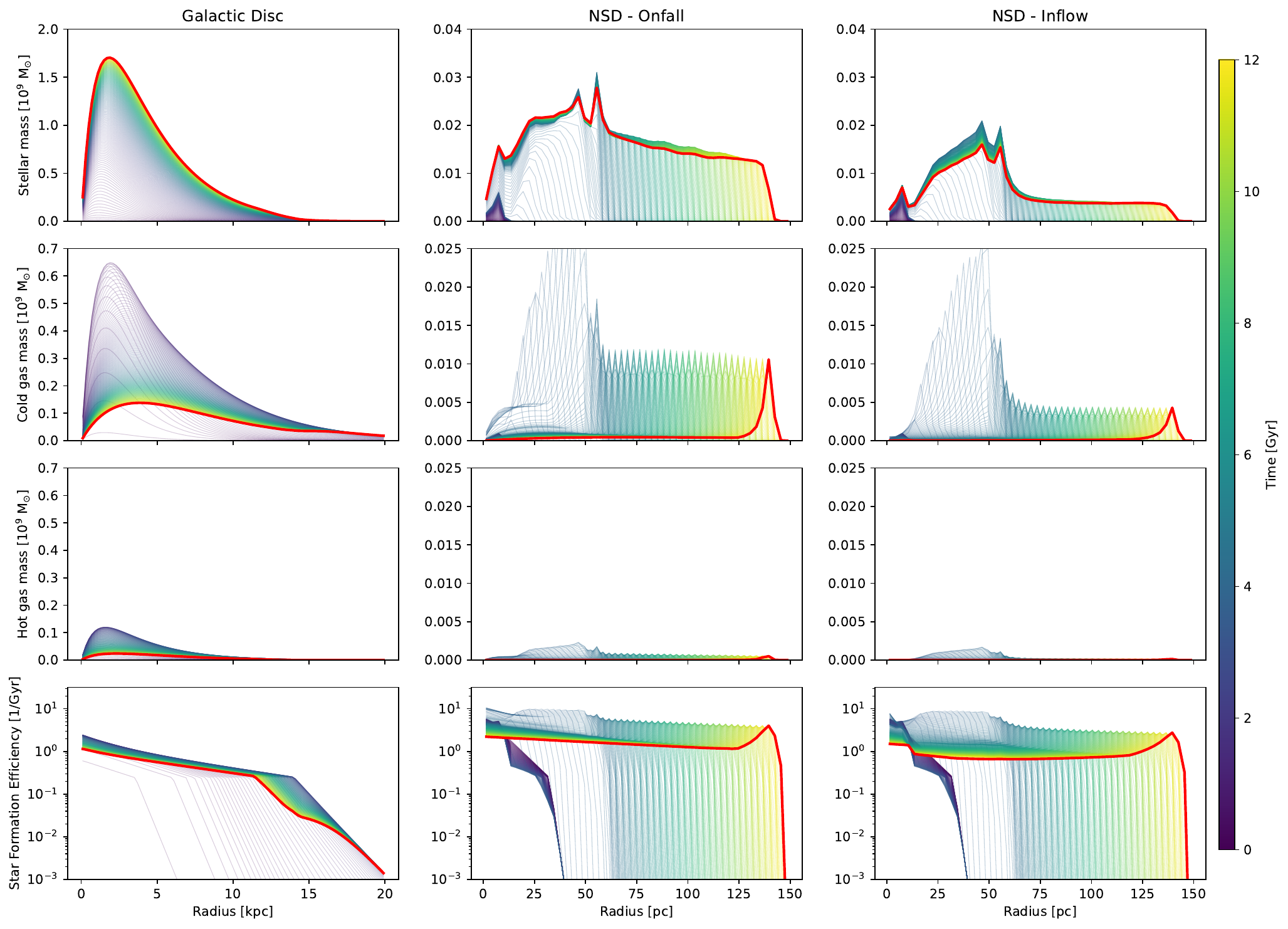}
     \caption{Stellar mass, mass of the cold and hot gas in each ring and star formation efficiency for the total galactic disc (left column) and the NSD for both of our accretion scenarios (right two columns). The thick red line in every panel shows the final mass/star formation efficiency profile.}
    \label{fig:gasmass}
    \end{figure*}

How the gas from the bar tips is accreted onto and distributed within the NSD is currently largely unconstrained. 

 Arguments that have been made about disc galaxies \citep[e.g. by][]{lacey_chemical_1985, marinacci_mode_2010, marasco_supernova-driven_2012, bilitewski_radial_2012} taking into account the angular momentum difference at the edge of a galaxy and predicting a balance of inflowing and onfalling gas at each radius cannot be easily transferred to the NSD: the angular momentum of the shocked gas coming from the tips of the bar as well as possible shielding of the NSD by the nuclear ring, preventing onfall, remain unclear. 

 It is also unclear how angular momentum is transferred within the NSD. \citet{sormani_fuelling_2023} measured the inflow velocity onto the NSD in NGC 1097, but this does not resolve the question of how much angular momentum this gas transfers to the NSD.

 Looking at the Milky Way, we can see a gas accretion feature onto the CMZ  around $ l \approx 1.3 ^{\circ}$ coinciding with the end of the bar dust lanes \citep{sormani_geometry_2019}. However, the inflow from the CMZ to the centre is much less clear, as bar-driven inflows become ineffective within the NSD \citep[e.g.][]{shlosman_fuelling_1990}.
  Some NSDs are known to exhibit substructures (particularly spirals) similar to large galactic discs \citep[e.g.][]{maciejewski_gas_2002,erwin_double-barred_2004}, and so both radial transport and stellar migration can be expected \citep[e.g.][]{shlosman_bars_1989}. Those substructures however seem to be more prevalent in larger discs.
  Further mechanisms driving gas flows are SN feedback \citep{tress_simulations_2020} and magnetic fields \citep{balbus_instability_1998, tress_magnetic_2024}. The expected inflow rate from those models is $\approx 0.02-0.03 \,{\rm M}_\odot \,{\rm yr}^{-1}$.  However, the full picture remains unclear \citep[for a discussion see also][]{henshaw_star_2023}.

To elucidate  the consequences and address our limited understanding of how material flows within NSDs, we consider two edge cases:
 
 The first one (the "onfall" scenario) achieves the surface density in eq.~\eqref{eq:surface_density} solely by onflow from the galactic disc. In other words, apart from a small blurring effect due to increased eccentricity of older stars, no stars or gas is exchanged between different rings of our model. 
 
 This is contrasted by our "inflow"  model, where the onfall is restricted to the area of the nuclear ring. The rest of the gas surface density is achieved by subsequent radial inflow through the NSD.
 
 We cap this inflow between neighbouring rings at each timestep at $10\%$ of the available mass. This means that it is possible that the prescribed surface density in eq.~\eqref{eq:surface_density} is not reached (a discussion of the impact of this parameter can be found in App.~\ref{sec:maxsteal}). The gas mass in the inflow scenario is on average half as high than the infall scenario, which  also leads to a on average star formation efficiency half as high as the the onflow scenario (compare Fig.~\ref{fig:gasmass}).

 Not knowing how much blurring or churning the NSD should have, we conduct this experiment with heavily suppressed radial migration in the NSD in either scenario. These can be measured once detailed stellar abundances and phase space information are known. (A test of the  impact of that parameter can be found in App.~\ref{fig:MixingStrength}.)
 
 We will later see that while the formation scenario strongly influences the radial abundance profile in the NSD, the global effect of the different gas parameters stays (mostly) the same.

The NSD has a comparably small circular velocity $v_\mathrm{c} {\sim} 125 \,{\rm km}\,{\rm s}^{-1}$, the system fills a quite small volume, and there is plentiful heating from the high rate of SNe. Together with the observations of outflow above the NSD, this leads to the assumption that the NSD cannot retain the majority of the hot ISM produced there. Here, we encode this by expelling $10\%$ of the present hot gas in each $10 \,{\rm Myr}$ time step (i.e. a residence timescale of $\sim 100 \,{\rm Myr}$ which is an order of magnitude smaller than the gas cooling time). 

One more possible gas accretion has to be considered: onfall from the galactic corona onto the NSD. As mentioned above, this accretion is poorly understood observationally and theoretically in the outer galactic disc, and there is currently no prediction if this happens at all in the very centre. The NSD is surrounded on both sides by the two Fermi bubbles (which may be driven by AGN eruptions and/or the NSD winds, first reported in \citealt{BlandHawthorn2003FermiDisc}, see further \citealt{su_giant_2010, dobler_fermi_2010, yang_unveiling_2018}). These likely shield the NSD against onfall. While they seem to be transient, it is likely that similar outflows prevail for most of the time above the NSD. Furthermore, the strongly star forming nuclear ring likely also drives outflows, preventing accretion of coronal gas. We hence do not include direct onfall from the CGM in our models. However, even if present, onfall from the Corona would reduce the overall metallicity, but is not expected to strongly affect abundance ratios.

As in the outer galactic disc, the NSD also loses a large fraction of its gas yields to the CGM. Due to its small vertical and radial extent and low angular momentum it is likely that this nuclear ejection fraction is significantly higher than the galactic disc ejection fraction, but even more difficult to directly measure. We nevertheless expect it to strongly affect the influence of the inflowing gas on the NSD. Therefore, we study its effect on the NSD abundance in our models in Sec.~\ref{sec:Results}.

\subsection{Summary of gas flows}
In summary we bracket the set of reasonable assumptions by our two main scenarios, setting the following assumptions on the gas flows:
\begin{itemize}
    \item No onfall from the CGM
    \item Similar hot/cold gas phase fraction in stellar yields as in the main galactic disc, but rapid gas loss from the hot phase. 
    \item Stellar yields in the bar region and gas from the bar tips are directly funnelled onto the NSD.    
\end{itemize}
Within the NSD, we either a) add the funnelled gas onto the outer ring and allow for an inflow sustaining the inner NSD (our favoured scenario), or b) add the funnelled gas directly onto each ring in the NSD without radial flows.

\section{Results}
    \label{sec:Results}
    \subsection{Mass Evolution of the NSD}
    
    \begin{figure}
        \centering
\includegraphics[width = 0.9\columnwidth]{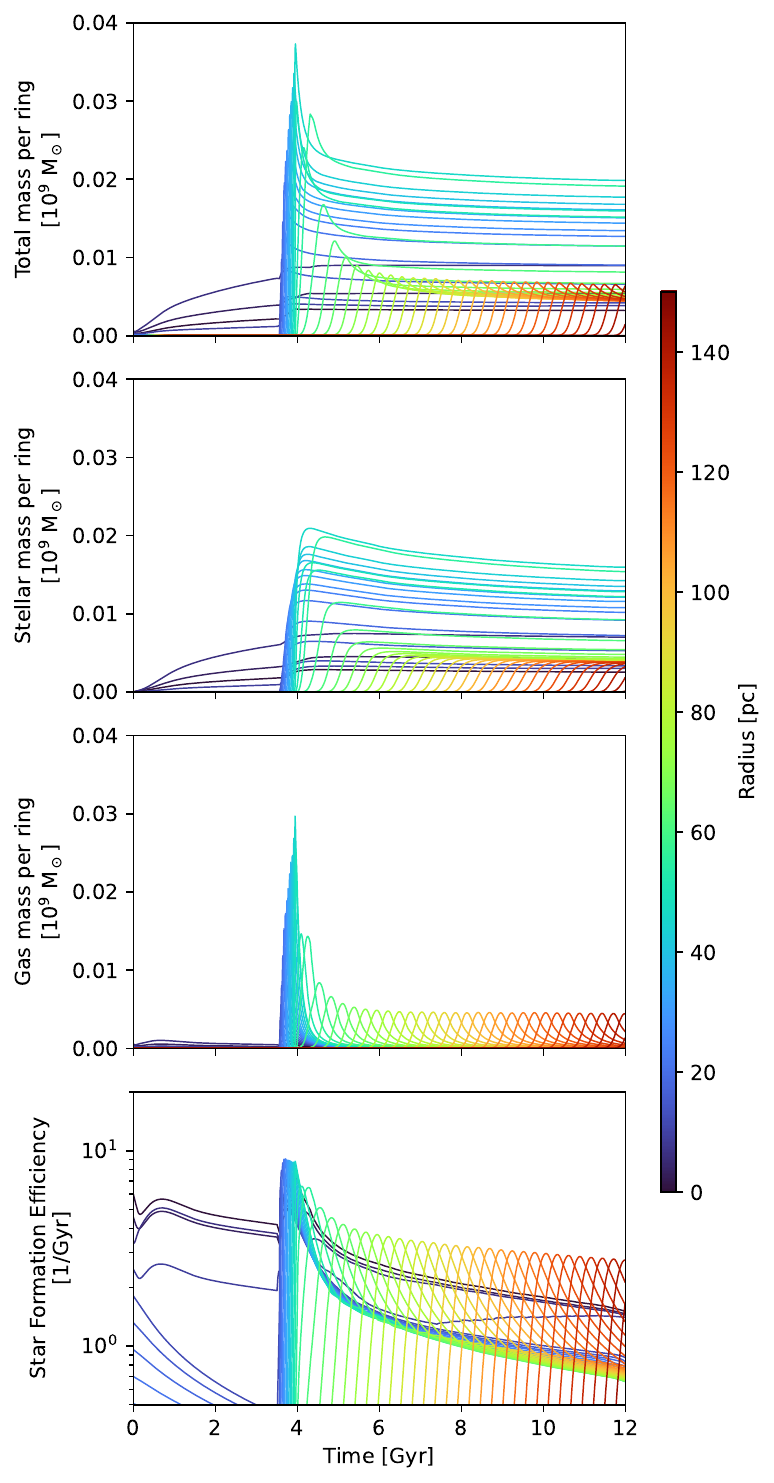}
    \caption{Total, stellar and gas mass per ring and star formation efficiency for the NSD in the inflow scenario. The changing colours from blue to red correspond to increasing radii.}
    \label{fig:gasmasstime}
    \end{figure}

    First we examine the mass profiles over time in the NSD. Fig.~\ref{fig:gasmass} shows the mass evolution (in stars and hot and cold gas) as well as star formation efficiency per ring for both the galactic disc as well as the NSD for both our accretion scenarios. The same data in terms of surface density can be found in Fig.~\ref{fig:gasmassSurfacedensity}. Additionally,  Fig.~\ref{fig:gasmasstime} shows the mass evolution over time for individual rings. In our set-up, the innermost ring receives a small amount of gas from the central galactic disc before bar formation, and blurring distributes some of this gas to the neighbouring rings. We note that this phase has virtually no effect on the results for the remainder of the simulation. During the initial bar formation around $t = 3.5 \,{\rm Gyr}$, the gas mass peaks in the central region of the NSD and drives a very strong and sustained star-formation burst with star formation efficiencies reaching up to $1/100 \,{\rm Myr}$. This leads to a rapid build-up of stellar mass in the inner rings. After this initial phase, the main star formation site in the NSD is the slowly expanding nuclear ring. While our model has continuous star formation, all our NSD models show a short decrease in the stellar mass after the initial star formation spike, when the high- and intermediate-mass stars from the burst die, followed by a slower subsequent build-up of mass. While the star-formation efficiency in the very central NSD still rivals that of the nuclear ring, this central region is small, and the ring dominates the star formation in the NSD, consistent with the observation of the ring in relatively young tip-RGB stars in \cite{schonrich_kinematic_2015}, while the overall mass density peaks clearly in the centre as demanded by infrared photometry \citep[e.g.][]{launhardt_nuclear_2002}. The final mass in the onfall model is $9.6 \times 10^{8} \,{\rm M}_\odot$, which is consistent with the estimate of $M_{\mathrm{NSD}} = 10.5 ^{+0.8}_{-0.7} \times 10^8 \,{\rm M}_\odot$ from \cite{sormani_self-consistent_2022}. The inflow model stays below this at $3.8 \times 10^8 \,{\rm M}_\odot$.

    \subsection{Gas Abundance Profile in the NSD}
    \label{sec:abundanceProfile}
    
\begin{figure*}
    \centering

    \includegraphics[width = 1.0\textwidth]{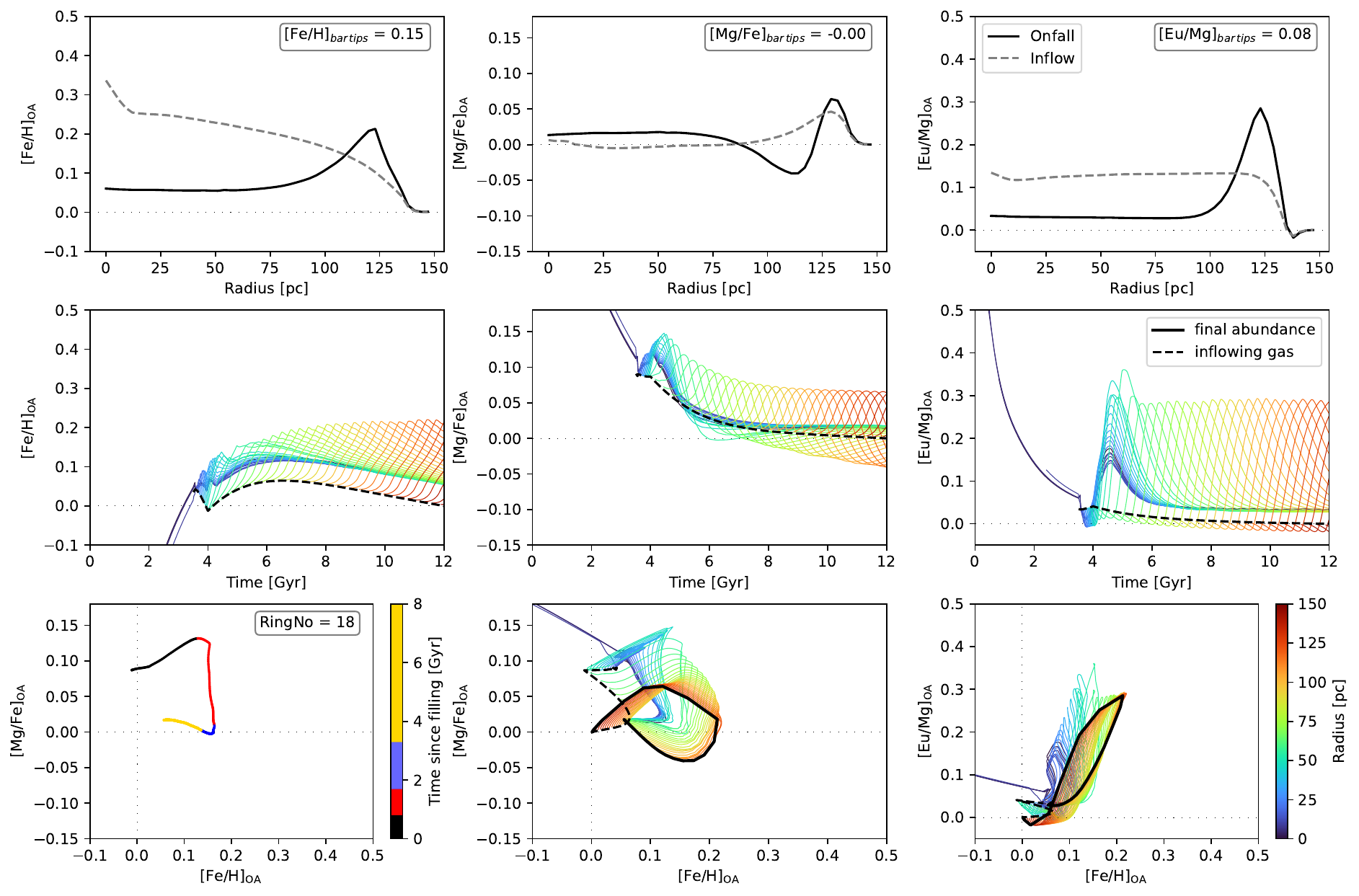}

    \caption{\textbf{Onfall scenario:} \textit{Top row}: Gas overabundance profiles in the NSD for \hbox{[Fe/H]$_\mathrm{OA}$}\xspace, $\hbox{[Mg/Fe]$_\mathrm{OA}$}\xspace$ and $\hbox{[Eu/Mg]$_\mathrm{OA}$}\xspace$ for our fiducial model in the onfall scenario (black solid line). The overabundance is calculated as the difference between the abundance at each radius in  the NSD and the abundance just outside the gas depleted bar region. For comparison, we also show the abundance profile of the competing inflow accretion history treated in Fig.~\ref{fig:gradientInflow} (gray dashed line). We also added the final abundances at the tips of the bar which are used for the calculation of the overabundance to the plots.
    \newline 
    \textit{Middle row:} General evolution of the elemental overabundance in the cold gas over time. The lines following each ring are plotted once the mass of the ring exceeds $4\times 10^6 \,{\rm M}_\odot$. The dashed line shows the abundance of the accreted gas over time.\newline
    \textit{Bottom row:} The panels in the middle and on the right give the relative abundance plane for magnesium over iron and europium over magnesium respectively. The final abundance is drawn in solid black and the abundance of the accreted gas as a dashed line. The leftmost panel shows again the magnesium abundance plane, but only for ring 18 at $R = 55.5\,{\rm pc}$ . The timescale is adjusted to start once this ring receives gas from the outer galaxy at $t = 4\,{\rm Gyr}$.}
    \label{fig:gradient}
\end{figure*}

\begin{figure*}
    \centering

        \includegraphics[width = 1.0\textwidth]{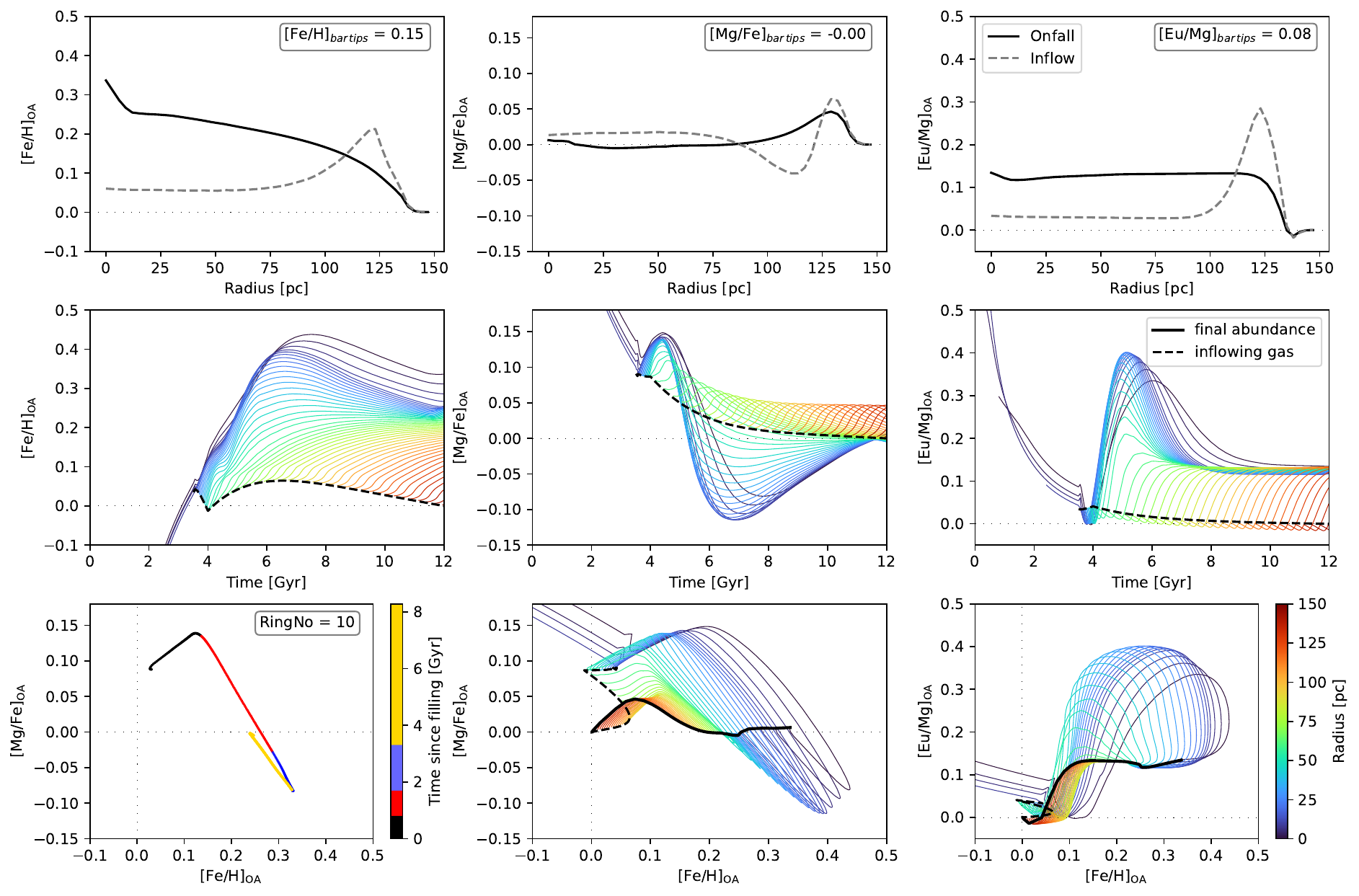}

        \caption{\textbf{Inflow scenario.} Otherwise following Fig.~\ref{fig:gradient}.}
    \label{fig:gradientInflow}
\end{figure*}
    
Before we vary individual parameters, we examine the gas abundance patterns predicted in the NSD at the end of the simulation for both the onfall and the inflow scenarios as shown in Fig.~\ref{fig:gradient} and Fig.~\ref{fig:gradientInflow}. (The corresponding plots in \hbox{[Eu/Fe]$_\mathrm{OA}$}\xspace can be found in Fig.~\ref{fig:metplaeEufe}, with an overall shape very similar to $\hbox{[Mg/Fe]$_\mathrm{OA}$}\xspace$, and an analogous explanation, while swapping in NSM for ccSN.)

We note that we present here results for the composition of the gas, which can be compared directly to observed gas phase and abundances of young stars. Modelling the older stellar populations would require an in-depth discussion of stellar re-distribution and selection function effects.

 The top row of Fig.~\ref{fig:gradient} shows today's (i.e. at $t = 12 \,{\rm Gyr}$) gas \textit{over}abundance (indicated by the subscript OA) of each NSD radius measured relative to the current bar tips near $R = 4 \,{\rm kpc}$.
 This means that a value of $0.1$ dex for example for $\hbox{[Fe/H]$_\mathrm{OA}$}\xspace$ in the top-left panel is not measured relative to the Sun, but that that radius in the NSD is $0.1$ dex (or $25\%$) more enriched in iron than the current star-forming gas at the bar tips from where the gas flows onto the NSD.
We make this choice to reduce the impact of nuisance parameters, like outer disc dynamics/history, and e.g. ejection rate parameters, which change the overall abundances, but say little about the NSD (see Sec.~\ref{sec:galacticDiscEjection} below for a detailed example of its impact). This relative measure should also be less biased in observations.

Comparing the top row of Fig.~\ref{fig:gradient} in both scenarios reveals strong qualitative differences, driven in parts by the different star formation patterns and most importantly, by different gas flow patterns. There are three competing influences on the abundance of each NSD ring in the inflow scenario (and two in the onfall scenario): i) the star-formation history and in-situ production of metals, ii) the time-dependent abundances of the funnelled gas from the outer disc, and in the inflow scenario also iii) the enrichment in the NSD outside the current radius. The overall abundances of the gas funnelled inwards (as seen in the dashed line in the middle row of Figs.~\ref{fig:gradient} and \ref{fig:gradientInflow}) are given by the overall chemical evolution of the galactic disc: a relatively swift initial rise in metallicity at high $\alpha$-element enrichment, which plateaus around the same time (of order 2-3 $\,{\rm Gyr}$) as SNIa lower the $\alpha$-abundances to solar values. Depending on its formation time, the NSD will experience the tail end of this phase of increasing metallicity and decreasing [$\alpha$/Fe] (the effects of a different bar formation time are shown in App.~\ref{sec:barformationepoch}). At later times, the change in infall composition is dominated by the radial abundance change due to the galactic metallicity gradient as the bar reaches larger radii, i.e. the metal-content of the funnelled gas is expected to slowly decline (shown in the dashed line in the middle left panel in Figure 4.2 at $t\gtrsim 6 \,{\rm Gyr}$).

The resulting patterns are best understood in two ways: i) by examining the trajectory in the abundance plane and hence the temporal evolution of single rings ii) by considering the relative star formation together with the change in gas (in)flows .

For the former we show the evolution of the  typical rings $18$ (and $10$ for the inflow scenario) in the abundance plane ($\hbox{[Mg/Fe]$_\mathrm{OA}$}\xspace$ vs. $\hbox{[Fe/H]$_\mathrm{OA}$}\xspace$) in the bottom left panel of Fig.~\ref{fig:gradient} (and Fig.~\ref{fig:gradientInflow} respectively). The trajectories are coloured by time since filling, i.e. since the nuclear ring has expanded to this radius. Each ring performs a clockwise loop in the abundance plane: With the initial star formation spike, both the overall metallicity and  $\hbox{[Mg/Fe]$_\mathrm{OA}$}\xspace$ quickly rise, as the ccSN enrich the star-forming nuclear ring with $\alpha$-elements and some iron. In the second phase (red), enrichment by SNIa kicks in, lowering the $\hbox{[Mg/Fe]$_\mathrm{OA}$}\xspace$ ratio while still raising the $\hbox{[Fe/H]$_\mathrm{OA}$}\xspace$ abundance. During this and the third phase (blue) the star-formation peak and strong gas accretion moves to the outside of the considered ring's radius and the importance of in-situ metal production declines. Here, the two scenarios behave quite differently. In the onfall scenario, the ring stays supplied from the bar tips. As these move slowly outward in this galaxy with a negative metallicity gradient, the overall metallicity of the funnelled gas keeps slowly declining, directly driving a corresponding decline in the ring metallicity. In the inflow scenario, all gas is delivered through the outer NSD with its current star formation spike, so the gas reaching the ring is pre-enriched by the increasingly large NSD on its outside, which compensates for some of the metallicity decline.

Let us now look at the time-evolution of all rings, shown in the middle row of both Fig.~\ref{fig:gradient} and \ref{fig:gradientInflow}.
In the onfall scenario, the overall trends of the funnelled gas can be clearly seen in the dashed line: the initial rise in $\hbox{[Fe/H]$_\mathrm{OA}$}\xspace$ is followed by a slow decline in the past $\sim 6 \,{\rm Gyr}$. Similarly $\hbox{[Mg/Fe]$_\mathrm{OA}$}\xspace$ shows a slow decline. As $\hbox{[Eu/Mg]$_\mathrm{OA}$}\xspace$ traces the relative importance of NSM and ccSN, the very slow decline in $\hbox{[Eu/Mg]$_\mathrm{OA}$}\xspace$ is a lingering effect of the larger NSM cold gas fraction ($f_{\mathrm{c,NSM}} > f_{\mathrm{c,ccSN}}$) which means that more of the ccSN yields from early galactic star formation remain locked up in the hot ISM phase and only gradually cool on a gigayear timescale.

Starting from these main trends, we see the changes at each radius after it is reached by the expanding nuclear ring: an initial rise in $\hbox{[Fe/H]$_\mathrm{OA}$}\xspace$, $\hbox{[Mg/Fe]$_\mathrm{OA}$}\xspace$, $\hbox{[Eu/Mg]$_\mathrm{OA}$}\xspace$ followed by a decline once the nuclear ring has passed. $\hbox{[Mg/Fe]$_\mathrm{OA}$}\xspace$ first rises sharply and is then taken slightly below the inflowing abundance when the ccSN burn out and SNIa from the initial star formation peak dominate. Due to the shifted timescales of NSM and ccSN, $\hbox{[Eu/Mg]$_\mathrm{OA}$}\xspace$ follows the inverse trend, exhibiting a dip first and then a rise when NSM dominate over ccSN, before eventually levelling out when the radius reaches equilibrium star formation. Again, due to the larger fraction of NSM yields being in the cold phase than ccSN, the rise is stronger than the dip. 
Similarly, we see in the bottom row of Figs.~\ref{fig:gradient} and \ref{fig:gradientInflow} that each radius initiates a clockwise loop around the main trend while it is passed by the nuclear ring.

\subsection{Influence of the Gas Inflow History}
     \label{sec:inflowVsOnfall}

It is instructive to analyse the differences between the onfall and inflow models. In the onfall scenario, rings are filled by directly funnelling gas from the bar tips. Hence, at every timestep, each ring receives gas with the same abundance, which is lower in metals than the processed gas in the NSD. By contrast, in the inflow model, fresh gas only reaches the nuclear ring and from there has to flow inwards through the NSD.

Let us now look at the observable abundance patterns vs. radius at present time which we show in the top row of Figs.~\ref{fig:gradient} and \ref{fig:gradientInflow}. In both models there is an overall heavy element enhancement as well as a slight $\hbox{[Mg/Fe]$_\mathrm{OA}$}\xspace$ and $\hbox{[Eu/Mg]$_\mathrm{OA}$}\xspace$ overabundance which is owed to the hot gas phase losses in the NSD that favours elements contributed by sources that inject more yields directly into the cold gas phase like ccSN or NSM (where $ f_{\mathrm{c,NSM}} > f_{\mathrm{c,ccSN}} \gg f_{\mathrm{c,SNIa}}$).

The most prominent feature, especially in the onfall scenario, is the overabundance peak of the nuclear ring at the outer edge of the NSD. As the star-forming ring slowly expands, the temporal sequence of star-burst induced chemical evolution loops \citep[see e.g.][]{colavitti_chemical_2008} becomes a function of radius: going inward from the current position, rings are in increasingly later stages of the burst. So, the outer side of the nuclear ring should be $\alpha$-enriched, the inner one iron enriched and [$\alpha$/Fe]$_\mathrm{OA}$ poor.

However, the main difference between the accretion models is the flow bringing metals from the NSD edge inwards in the inflow scenario. Consequently, the inflow model shows a much larger accumulation of heavy elements, as the inflowing gas gets successively enriched on its way to the centre, leading to a large negative radial $\hbox{[Fe/H]$_\mathrm{OA}$}\xspace$ gradient.
A more subtle difference is the dip in $\hbox{[Mg/Fe]$_\mathrm{OA}$}\xspace$ just inside the nuclear ring in the onfall model. This dip is filled by the high-$\hbox{[Mg/Fe]}\xspace$ material flowing inwards from the ring in the inflow model. Similarly, the high $\hbox{[Eu/Mg]$_\mathrm{OA}$}\xspace$ peak seen in the onfall model and created by the shifted onset of NSM and ccSN distributes across the whole NSD in the inflow model.

    \subsection{Varying the Parameters}
    \label{sec:EffectsOfTheMainGasParameters}
    \setlength{\tabcolsep}{1pt}

\begin{figure*}
        \centering      
        \begin{tabular}{m{0.5cm}m{0.235\textwidth}m{0.22\textwidth}m{0.22\textwidth}m{0.22\textwidth}p{.01cm}p{0.5cm}}
                \multicolumn{7}{c}{\textbf{Onfall}}\\
                &\makecell{\footnotesize total\\ \footnotesize abundance} \hspace{.5cm} & \makecell{\footnotesize overabundance} \hspace{.5cm} & \makecell{\footnotesize total \\  \footnotesize - median \\\footnotesize abundance} \hspace{.5cm}&\makecell{\footnotesize overabundance  \\\footnotesize - median \\\footnotesize overabundance}\hspace{.5cm}&&\\
                 \parbox[t]{20mm}{\multirow{3}{*}{\rotatebox[origin=l]{90}{\footnotesize \textbf{Galactic Disc Ejection Fraction}\hspace{1cm}}}}&
                 
         \includegraphics[width = 0.235\textwidth, trim = 15 22 36 25, clip, page = 5]{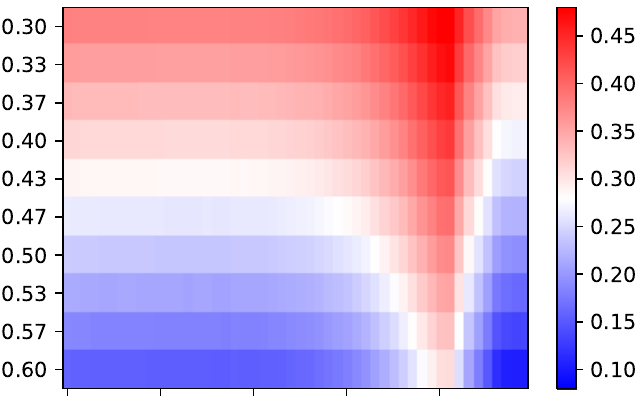}&  
         \includegraphics[width = 0.22\textwidth,  trim = 40 22 30 25, clip, page = 5]{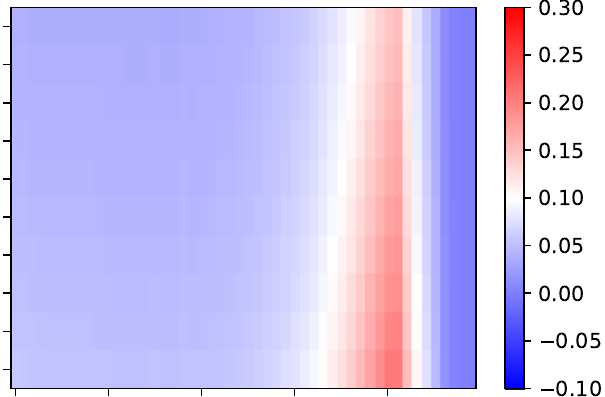}  &
         \includegraphics[width = 0.22\textwidth, trim = 40 22 30 25, clip, page = 5]{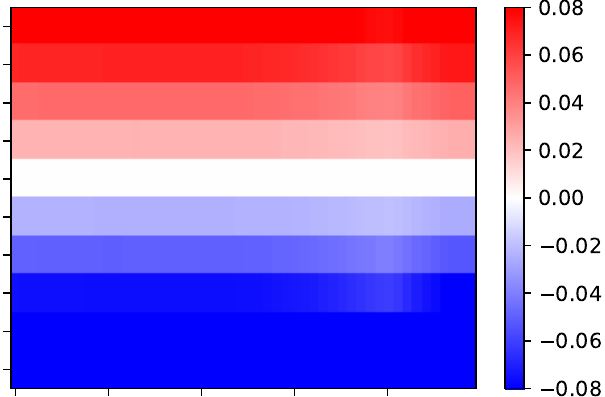}   &
         \includegraphics[width = 0.22\textwidth, trim = 40 22 30 25, clip, page = 5]{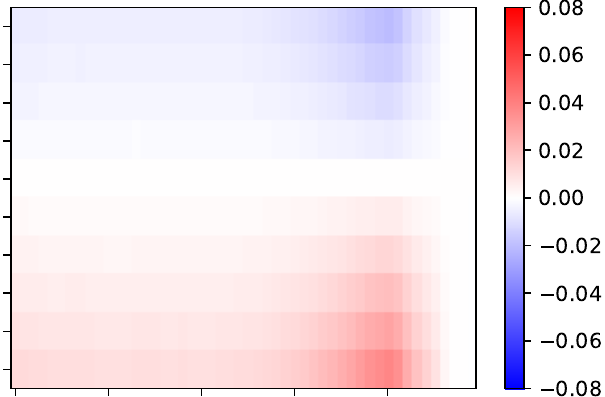}  
         & &\parbox[t]{20mm}{\multirow{3}{*}{\rotatebox[origin=l]{90}{\footnotesize \hspace{0cm}[Eu/Fe]\hspace{1.8cm} [Mg/Fe] \hspace{1.8cm} [Fe/H] \hspace{-0.5cm} }}} \\
         &
         \includegraphics[width = 0.236\textwidth, trim = 12 22 32 25, clip, page = 5]{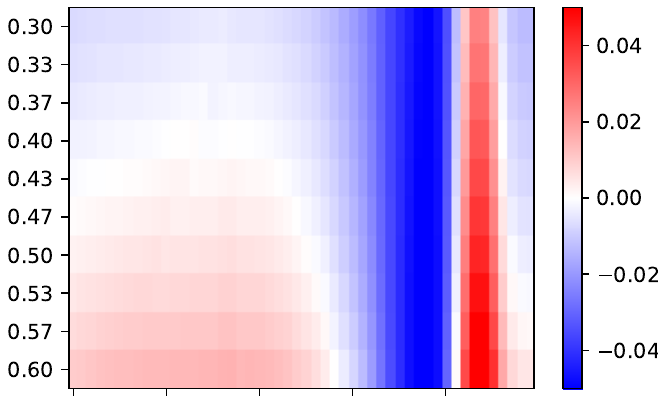}& 
         \includegraphics[width = 0.22\textwidth, trim = 40 22 30 25, clip, page = 5]{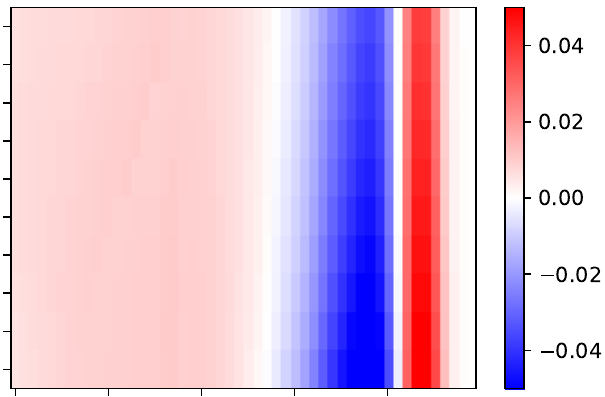}  &
         \includegraphics[width = 0.22\textwidth, trim = 40 22 30 25, clip, page = 5]{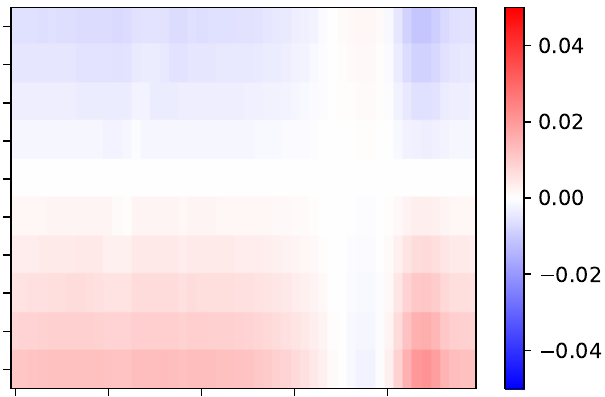} &
         \includegraphics[width = 0.22\textwidth, trim = 40 22 30 25, clip, page = 5]{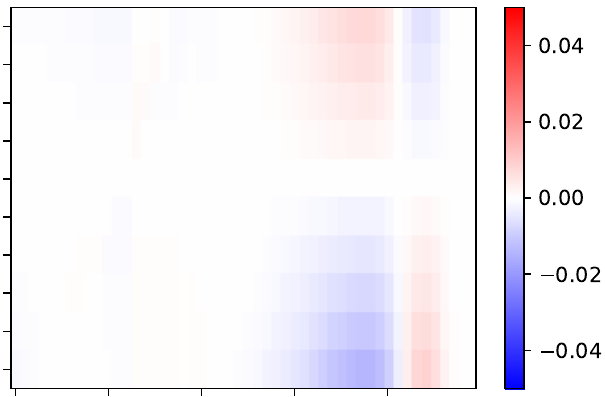} 
         & &\\
         & 
         \includegraphics[width = 0.235\textwidth, trim = 12 10 36 25, clip, page = 5]{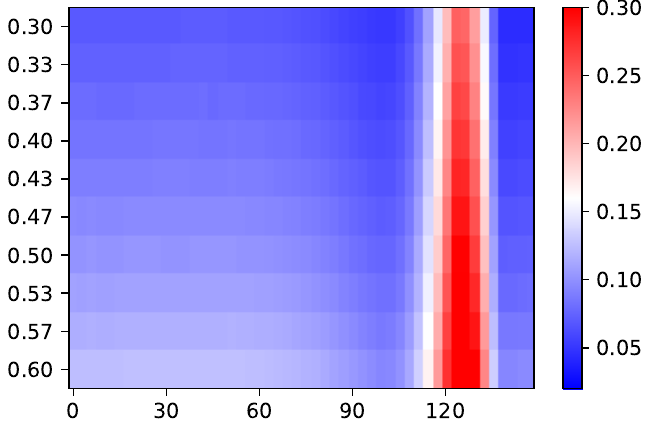}& 
         \includegraphics[width = 0.22\textwidth, trim = 40 10 30 25, clip, page = 5]{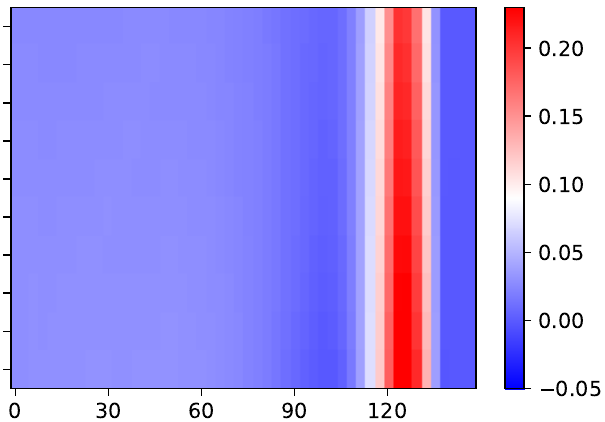}  &
         \includegraphics[width = 0.22\textwidth, trim = 40 10 30 25, clip, page = 5]{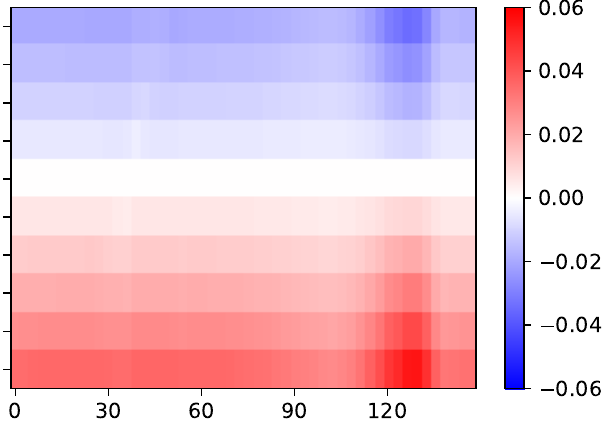} &
         \includegraphics[width = 0.22\textwidth, trim = 40 10 30 25, clip, page = 5]{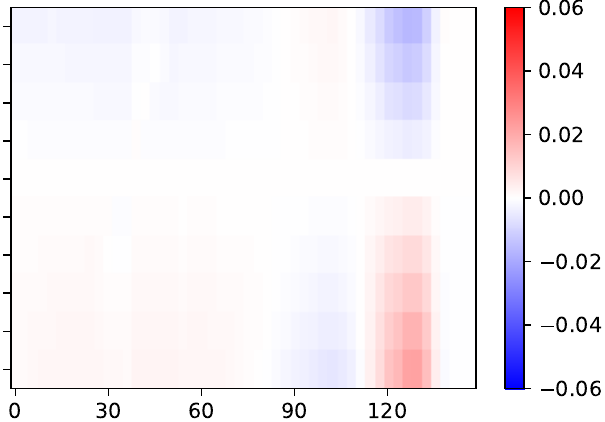} 
         &&\\
         \multicolumn{7}{c}{\footnotesize \textbf{Radius[pc]}}\\
        \hline\\
        \multicolumn{7}{c}{\textbf{Inflow}}\\
        
        &\makecell{\footnotesize total\\ \footnotesize abundance} \hspace{.5cm} & \makecell{\footnotesize overabundance} \hspace{.5cm} & \makecell{\footnotesize total \\  \footnotesize - median \\\footnotesize abundance} \hspace{.5cm}&\makecell{\footnotesize overabundance  \\\footnotesize - median \\\footnotesize overabundance}\hspace{.5cm}&&\\

        \parbox[t]{20mm}{\multirow{3}{*}{\rotatebox[origin=l]{90}{\footnotesize\textbf{Galactic Disc Ejection Fraction}\hspace{1cm}}}}&

         \includegraphics[width = 0.235\textwidth, trim = 15 22 36 25, clip, page = 5]{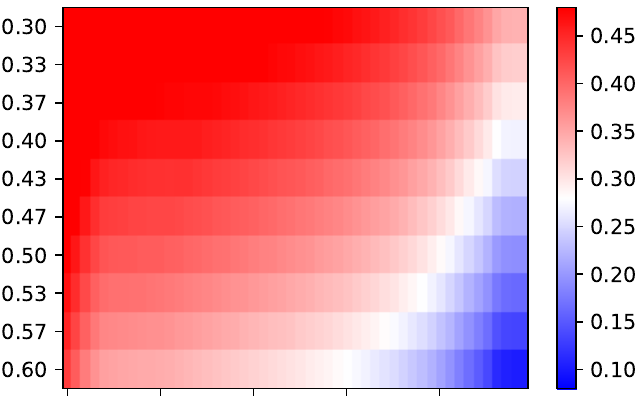}&  
         \includegraphics[width = 0.22\textwidth,  trim = 40 22 30 25, clip, page = 5]{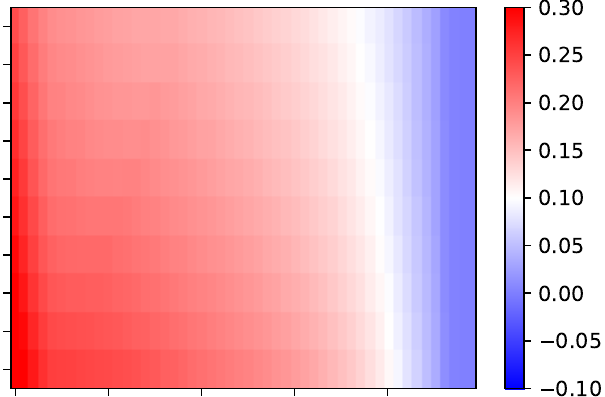}  &
         \includegraphics[width = 0.22\textwidth, trim = 40 22 30 25, clip, page = 5]{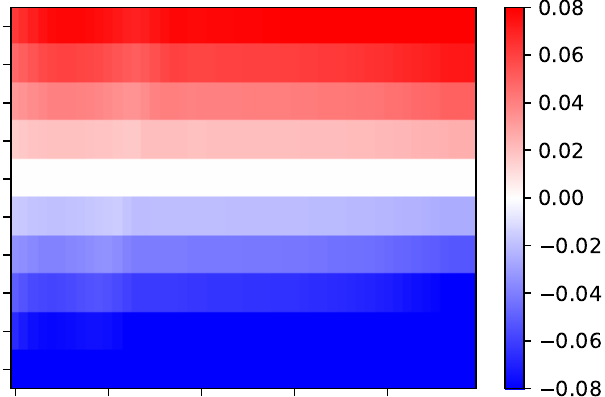}   &
         \includegraphics[width = 0.22\textwidth, trim = 40 22 30 25, clip, page = 5]{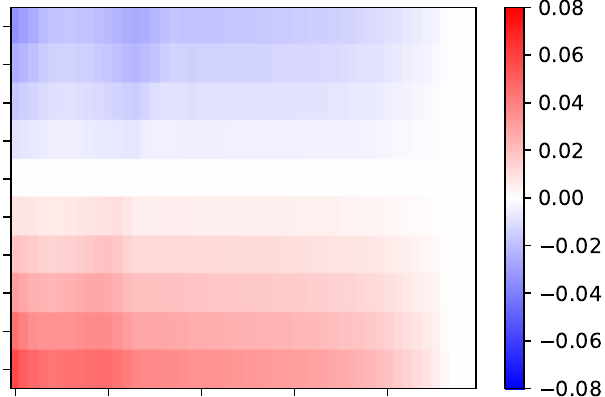}  
         & &\parbox[t]{20mm}{\multirow{3}{*}{\rotatebox[origin=l]{90}{\footnotesize \hspace{0cm}[Eu/Fe]\hspace{1.8cm} [Mg/Fe] \hspace{1.8cm} [Fe/H] \hspace{-0.5cm} }}} \\

         &
         \includegraphics[width = 0.236\textwidth, trim = 12 22 32 25, clip, page = 5]{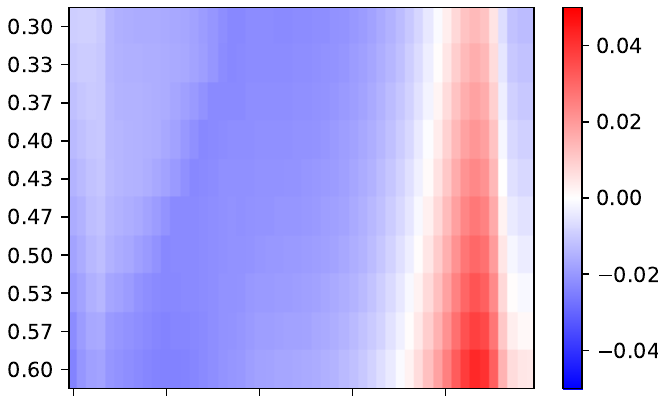}& 
         \includegraphics[width = 0.22\textwidth, trim = 40 22 30 25, clip, page = 5]{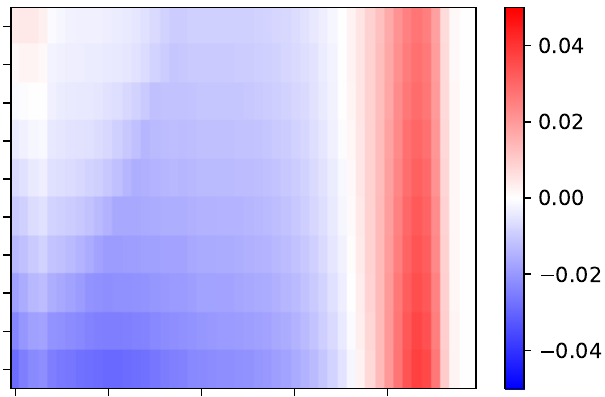}  &
         \includegraphics[width = 0.22\textwidth, trim = 40 22 30 25, clip, page = 5]{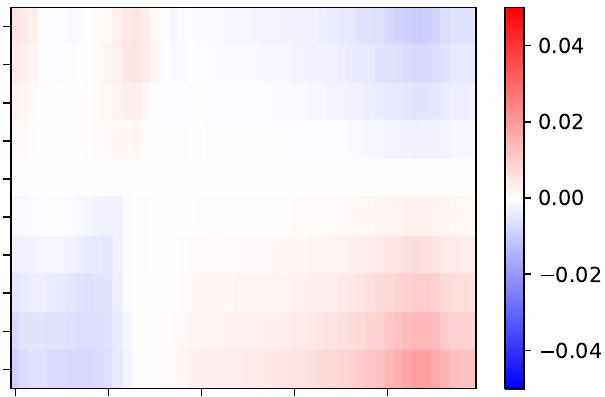} &
         \includegraphics[width = 0.22\textwidth, trim = 40 22 30 25, clip, page = 5]{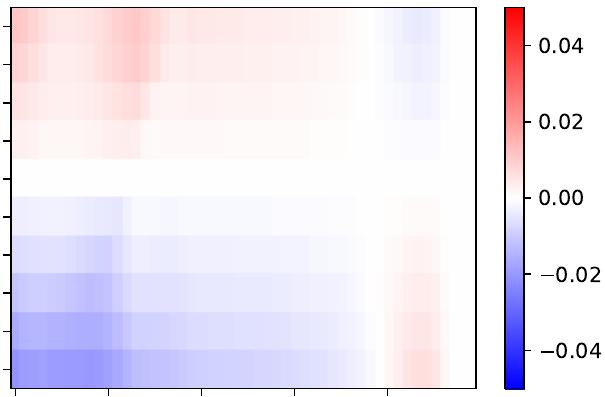} 
         &&\\
         & 
         \includegraphics[width = 0.235\textwidth, trim = 12 10 36 25, clip, page = 5]{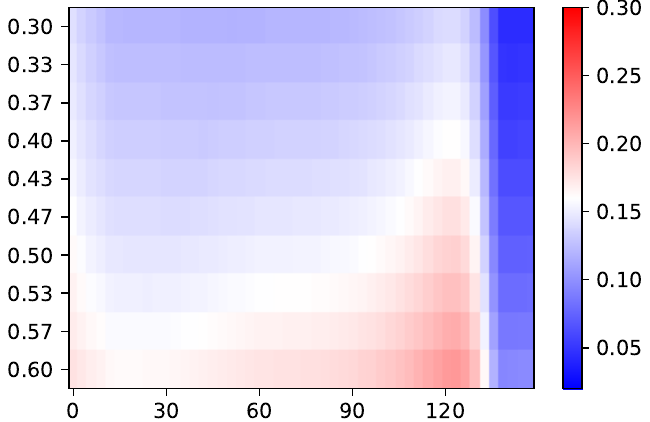}& 
         \includegraphics[width = 0.22\textwidth, trim = 40 10 30 25, clip, page = 5]{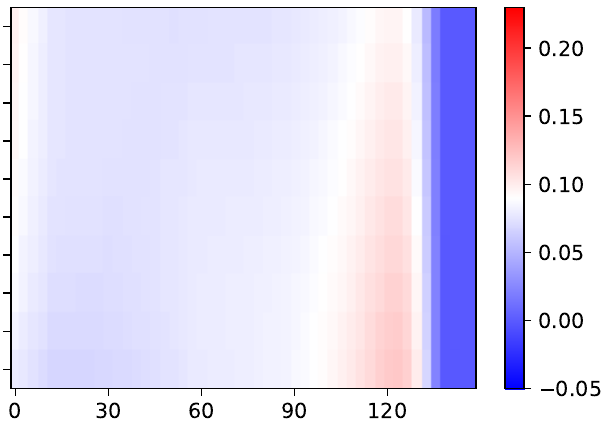}  &
         \includegraphics[width = 0.22\textwidth, trim = 40 10 30 25, clip, page = 5]{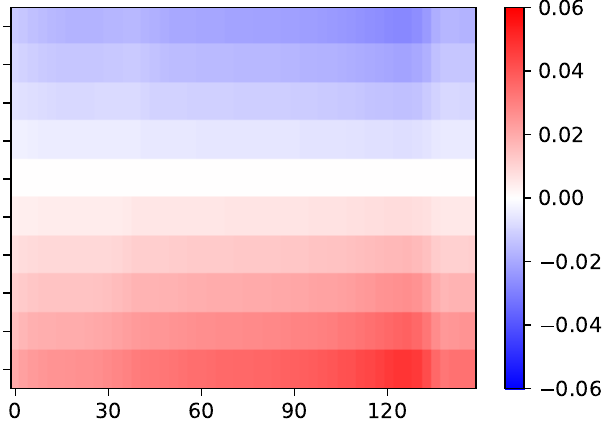} &
         \includegraphics[width = 0.22\textwidth, trim = 40 10 30 25, clip, page = 5]{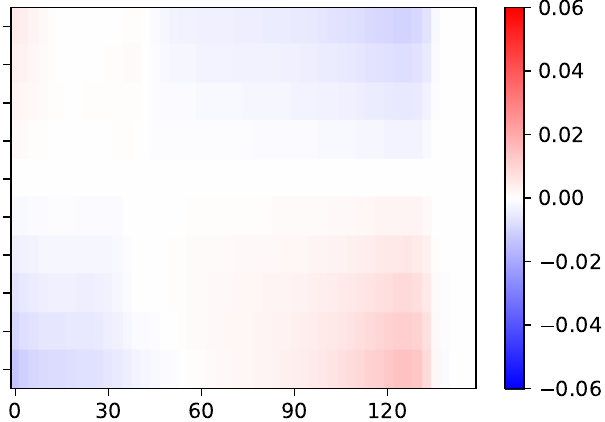} 
         &&\\
         \multicolumn{7}{c}{\footnotesize\textbf{Radius[pc]}}\\

        \end{tabular}
        \caption{Effects of the galactic disc ejection fraction on the [Fe/H], [Mg/Fe], [Eu/Fe] gas (over) abundance in the NSD. Columns from the left: total abundance, overabundance over the final abundance at the tips of the galactic bar at $4\,{\rm kpc}$, and then total and total overabundance minus the median model profile. While the two columns on the left do not have the same colour baseline, they share the same total abundance delta. The two right columns are row-wise on the same scale.}
        \label{fig:ejectglobAbundance}
    \end{figure*}

After the discussing the main abundance features in the last sections, we turn to varying the parameters guiding the global gas balances. Here, we mainly focus on the two ejection fractions, i.e. the fraction of yields that is lost to the CGM in the galactic and the nuclear disc. 

\subsubsection{Galactic Disc Ejection Fraction}
\label{sec:galacticDiscEjection}
The ejection fraction to the CGM in a galaxy is a simplification for the observed mass loss of processed gas from starforming discs. We remind the reader also of the analytical solution for the equilibrium abundance of some element $j$, $Z_{j, eq}$, in an open box model,
\begin{equation}
Z_{j, eq} = Z_{j,y} \frac{(1- e_s)r}{1+ \eta},
\end{equation}
where $Z_{j,y}$ is the abundance of the stellar yields, $e_s$ the ejection fraction of yields from the stars into the CGM, $\eta$ the outflow efficiency or ablation factor and $r$ the recycling factor, i.e. the fraction of stellar yields that does not stay indefinitely locked up in low mass stars. 
In the real world, when the galactic disc looses a higher fraction of its yields with increasig $e_s$, the lost gas mass is compensated by more metal-poor inflowing gas, lowering the metallicity similiar to the ablation term $\eta$. Since the main galactic disc feeds the nuclear disc, lowering its metallicity also reduces elemental abundances in the NSD. 

We can trace this effect in Fig.~\ref{fig:ejectglobAbundance}, where we show the final values of $\hbox{[Fe/H]}\xspace$, $\hbox{[Mg/Fe]}\xspace$, and $\hbox{[Eu/Fe]}\xspace$ (note that we  trace $\hbox{[Eu/Fe]}\xspace$ here not $\hbox{[Eu/Mg]}\xspace$ as above) as a function of radius in the nuclear disc while varying the galactic disc ejection fraction between 30 and $60\%$ (declining along the y-axis),  keeping the other parameters at their fiducial values from table \ref{tab:simparam_bestfit}. As before, we compare the onfall and inflow models in the upper and lower parts of the figure.  
In the plot, we see in the left two columns the resulting abundance profiles (relative to the Sun and the tips of the bar respectively), whereas the right-hand side of the plot (columns 3 and 4) shows the (over)abundances relative to the median parametrisation - hence a white line at loss fraction $0.43$.\footnote{Note that the median models here are not fully equivalent to the fiducial model presented in Tab.~\ref{tab:simparam_bestfit} as the interval in which we vary the GCE parameters are not necessarily centred around the fiducial value. Hence, for example in case of varying the galactic disc ejection fraction, all other parameters correspond to their fiducial values from Tab.~\ref{tab:simparam_bestfit}, while the galactic disc ejection fraction has the median value within the tested interval of [0.3,0.6] which is 0.43.}

We first focus on the onfall model. The galactic disc ejection fraction strongly affects the overall metallicity, seen in $\hbox{[Fe/H]}\xspace$ (first row). Additionally, also the abundance ratios $\hbox{[Mg/Fe]}\xspace$ and $\hbox{[Eu/Fe]}\xspace$ show a weak effect. 
However, due to different hot gas fractions between ccSN/NSM and SNIa, iron is most strongly affected as its retention in the NSD is most strongly suppressed. 
This seems at first concerning, since the galactic disc ejection fraction is not well-constrained in chemical evolution. However, this is greatly helped by using the relative overabundances relative to the bar tips (2nd/4th column in Fig. \ref{fig:ejectglobAbundance}), which reduces the effect by about an order of magnitude. In the onfall scenario, the remaining change is quite straight-forward. Focusing on the 4\textsuperscript{th} column, which concentrates on the changes between models by subtracting the abundance profile of the median model, one can see that  increasing the galactic disc ejection fraction lowers the overall baseline abundance. Hence, $\hbox{[Fe/H]$_\mathrm{OA}$}\xspace$ overabundances in the NSD come out slightly higher. 
On the abundance ratios, the same effect exacerbates slightly the overabundance pattern at smaller ejection fractions: We see a pronounced lowering of the $\hbox{[Mg/Fe]}\xspace$ ratio and (due to the smaller difference in timescales between NSM and SNIa than ccSN and SNIa) a milder lowering of $\hbox{[Eu/Fe]}\xspace$ directly within the nuclear ring where the star formation rate is dropping steeply as the ring expands outwards. This means higher current iron production, and so the lower baseline metallicity lowers $\hbox{[Mg/Fe]}\xspace$ ratios behind the ring. At the same time, the relative overabundances \hbox{[Mg/Fe]$_\mathrm{OA}$}\xspace and \hbox{[Eu/Fe]$_\mathrm{OA}$}\xspace at the location of the star forming nuclear ring are stronger at higher ejection fractions, again because in this case, the current star formation burst on the ring has a stronger effect on the gas abundances.

\begin{figure*}
    \centering
    \begin{tabular}{m{0.5cm}m{0.235\textwidth}m{0.22\textwidth}m{0.22\textwidth}m{0.22\textwidth}p{.01cm}p{0.5cm}}
    \multicolumn{7}{c}{\textbf{Onfall}}\\
    &\makecell{\footnotesize total\\ \footnotesize abundance} \hspace{.5cm} & \makecell{\footnotesize overabundance} \hspace{.5cm} & \makecell{\footnotesize total \\  \footnotesize - median \\\footnotesize abundance} \hspace{.5cm}&\makecell{\footnotesize overabundance  \\\footnotesize - median \\\footnotesize overabundance}\hspace{.5cm}&&\\
     \parbox[t]{20mm}{\multirow{3}{*}{\rotatebox[origin=l]{90}{\footnotesize\textbf{NSD Ejection Fraction}\hspace{1cm}}}}&
     
     \includegraphics[width = 0.235\textwidth, trim = 15 22 36 25, clip, page = 6]{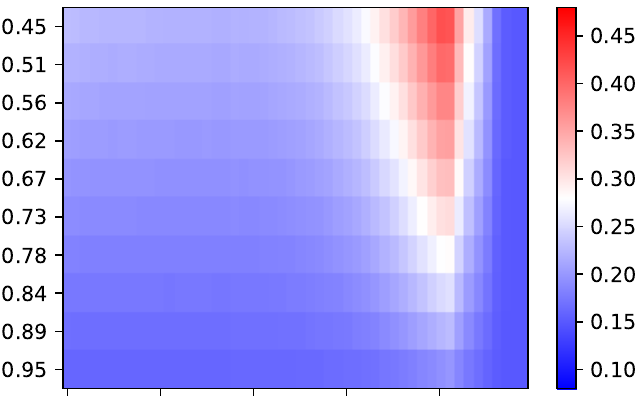}&  
     \includegraphics[width = 0.22\textwidth,  trim = 40 22 30 25, clip, page = 6]{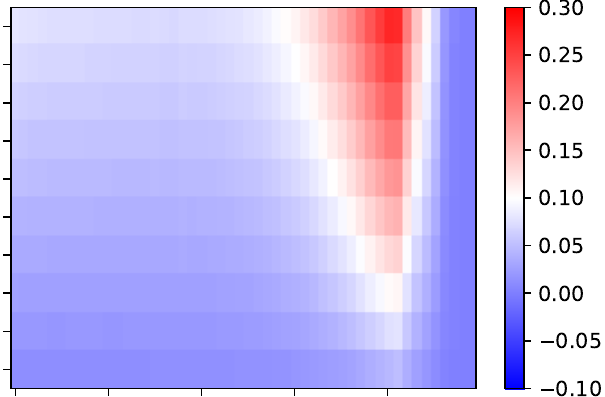}  &
     \includegraphics[width = 0.22\textwidth, trim = 40 22 30 25, clip, page = 6]{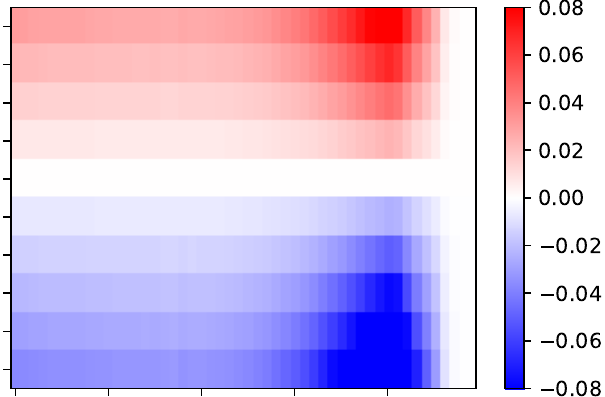}   &
     \includegraphics[width = 0.22\textwidth, trim = 40 22 30 25, clip, page = 6]{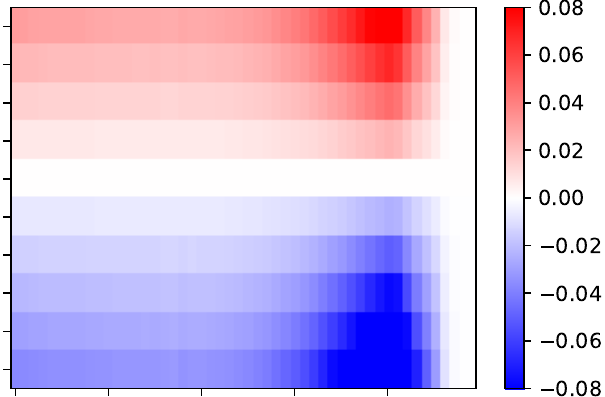}  
     & &\parbox[t]{20mm}{\multirow{3}{*}{\rotatebox[origin=l]{90}{\footnotesize \hspace{0cm}[Eu/Fe]\hspace{1.8cm} [Mg/Fe] \hspace{1.8cm} [Fe/H] \hspace{-0.5cm} }}} \\
     &
     \includegraphics[width = 0.24\textwidth, trim = 12 22 32 25, clip, page = 6]{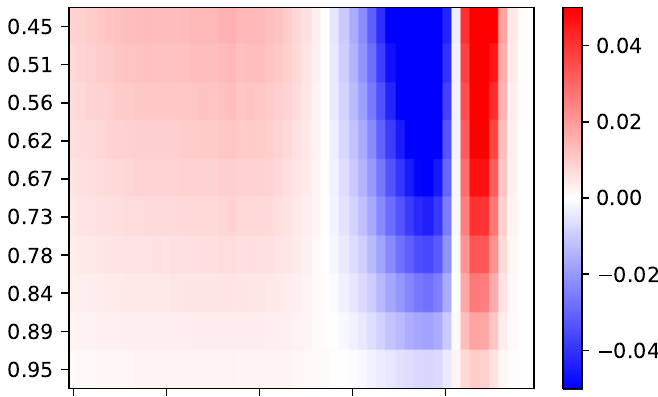}& 
     \includegraphics[width = 0.22\textwidth, trim = 40 22 30 25, clip, page = 6]{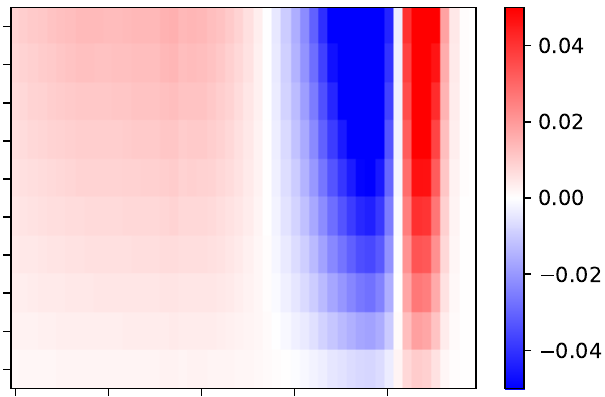}  &
     \includegraphics[width = 0.22\textwidth, trim = 40 22 30 25, clip, page = 6]{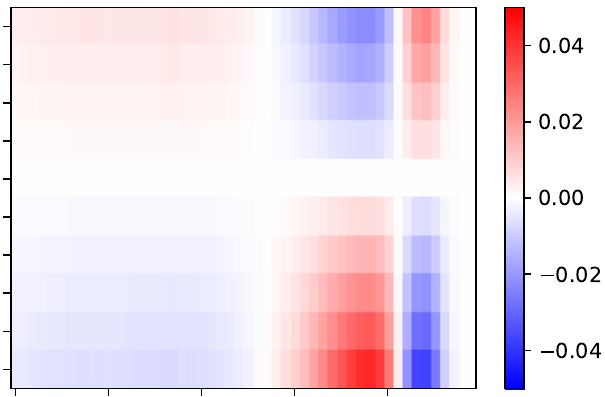} &
     \includegraphics[width = 0.22\textwidth, trim = 40 22 30 25, clip, page = 6]{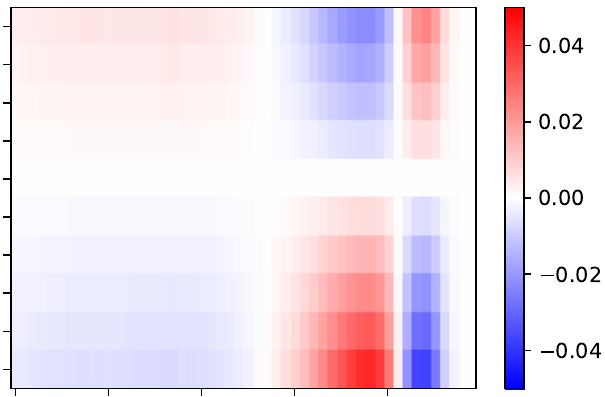} 
     && \\
     & 
     \includegraphics[width = 0.235\textwidth, trim = 12 10 36 25, clip, page = 6]{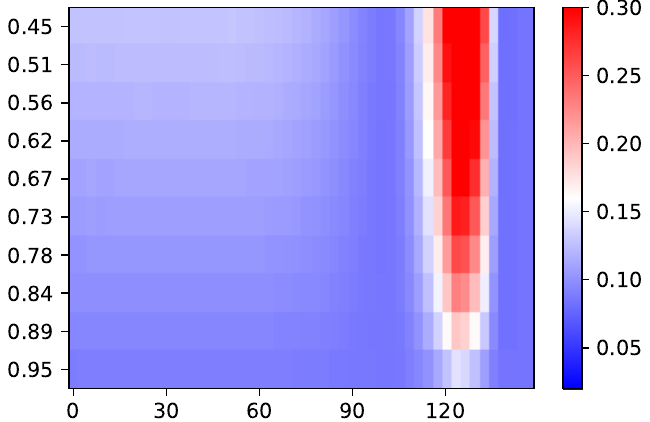}& 
     \includegraphics[width = 0.22\textwidth, trim = 40 10 30 25, clip, page = 6]{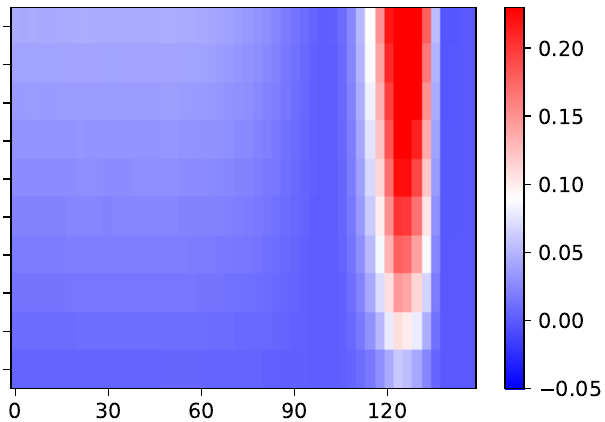}  &
     \includegraphics[width = 0.22\textwidth, trim = 40 10 30 25, clip, page = 6]{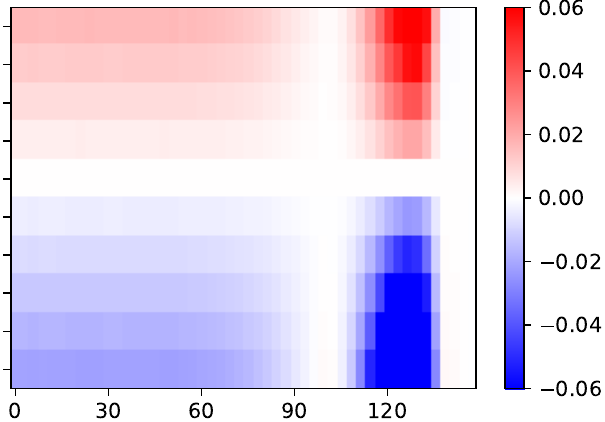} &
     \includegraphics[width = 0.22\textwidth, trim = 40 10 30 25, clip, page = 6]{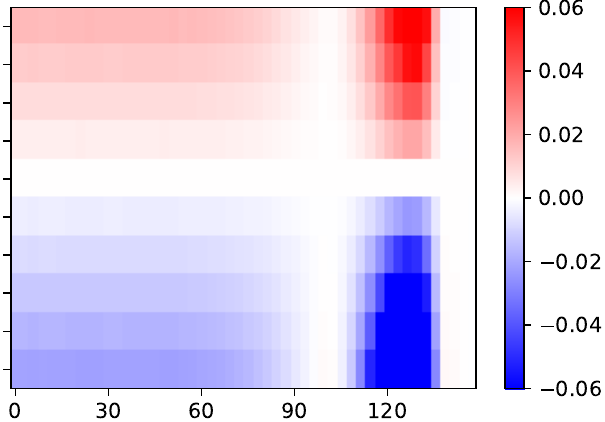} 
     &&\\
     \multicolumn{7}{c}{\footnotesize\textbf{Radius[pc]}}\\
    \hline\\
    \multicolumn{7}{c}{\textbf{Inflow}}\\
    &\makecell{\footnotesize total\\ \footnotesize abundance} \hspace{.5cm} & \makecell{\footnotesize overabundance} \hspace{.5cm} & \makecell{\footnotesize total \\  \footnotesize - median \\\footnotesize abundance} \hspace{.5cm}&\makecell{\footnotesize overabundance  \\\footnotesize - median \\\footnotesize overabundance}\hspace{.5cm}&&\\
     \parbox[t]{20mm}{\multirow{3}{*}{\rotatebox[origin=l]{90}{\footnotesize\textbf{NSD Ejection Fraction}\hspace{1cm}}}}&
     
     \includegraphics[width = 0.235\textwidth, trim = 15 22 36 25, clip, page = 6]{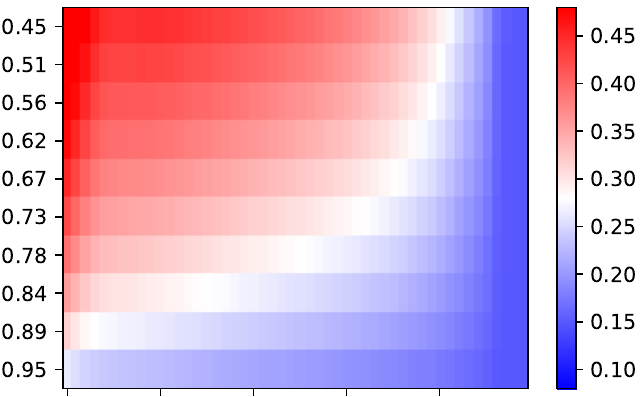}&  
     \includegraphics[width = 0.22\textwidth,  trim = 40 22 30 25, clip, page = 6]{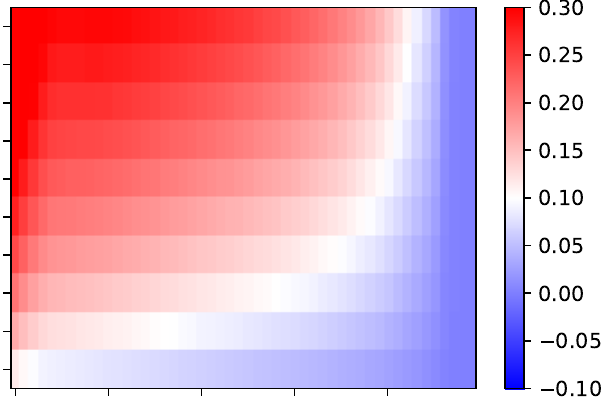}  &
     \includegraphics[width = 0.22\textwidth, trim = 40 22 30 25, clip, page = 6]{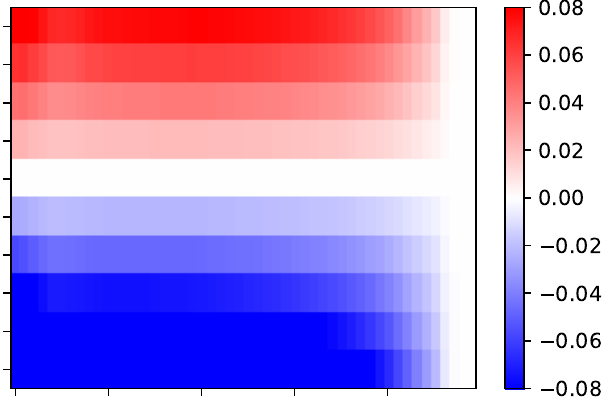}   &
     \includegraphics[width = 0.22\textwidth, trim = 40 22 30 25, clip, page = 6]{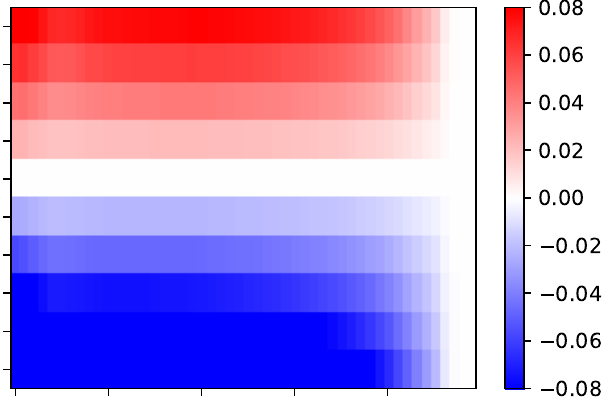}  
     & &\parbox[t]{20mm}{\multirow{3}{*}{\rotatebox[origin=l]{90}{\footnotesize \hspace{0cm}[Eu/Fe]\hspace{1.8cm} [Mg/Fe] \hspace{1.8cm} [Fe/H] \hspace{-0.5cm} }}} \\

     &
     \includegraphics[width = 0.24\textwidth, trim = 12 22 32 25, clip, page = 6]{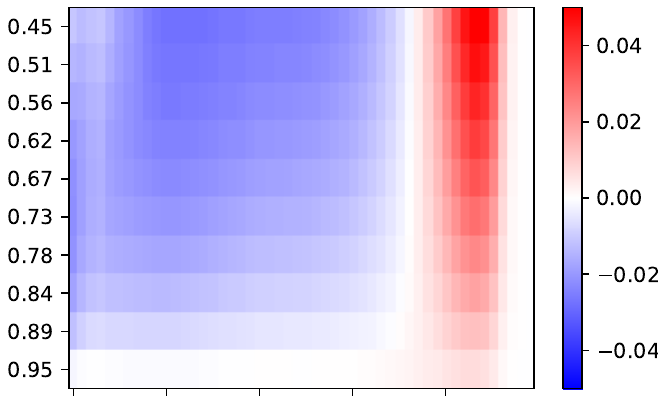}& 
     \includegraphics[width = 0.22\textwidth, trim = 40 22 30 25, clip, page = 6]{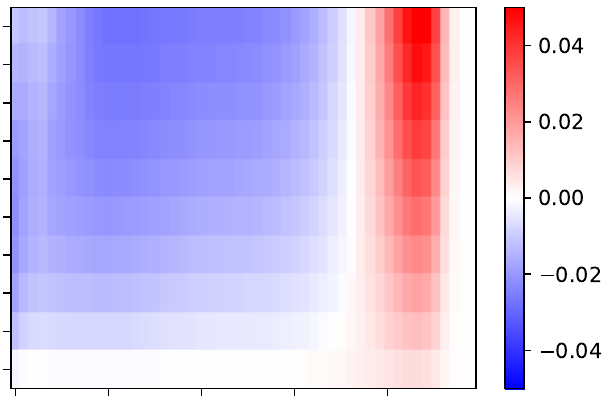}  &
     \includegraphics[width = 0.22\textwidth, trim = 40 22 30 25, clip, page = 6]{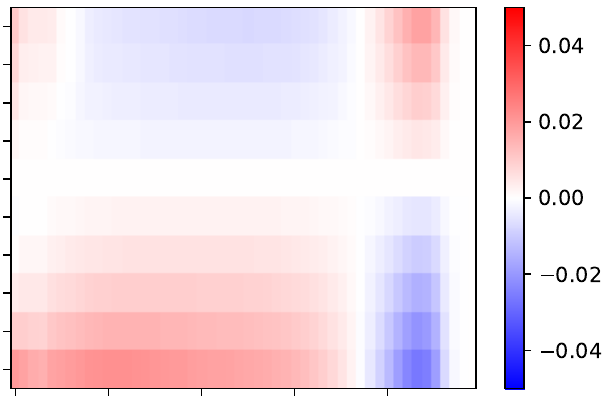} &
     \includegraphics[width = 0.22\textwidth, trim = 40 22 30 25, clip, page = 6]{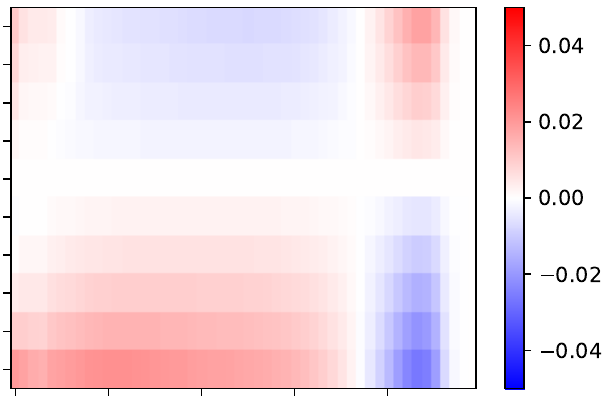} 
     &&\\
     & 
     \includegraphics[width = 0.235\textwidth, trim = 12 10 36 25, clip, page = 6]{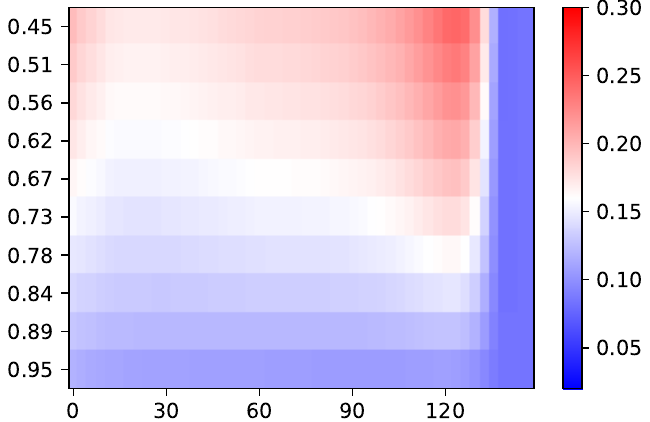}& 
     \includegraphics[width = 0.22\textwidth, trim = 40 10 30 25, clip, page = 6]{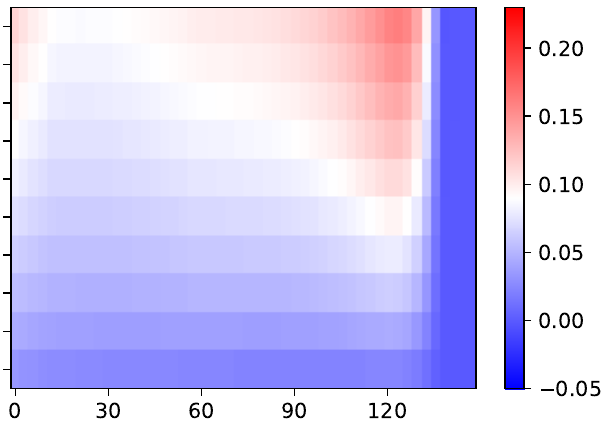}  &
     \includegraphics[width = 0.22\textwidth, trim = 40 10 30 25, clip, page = 6]{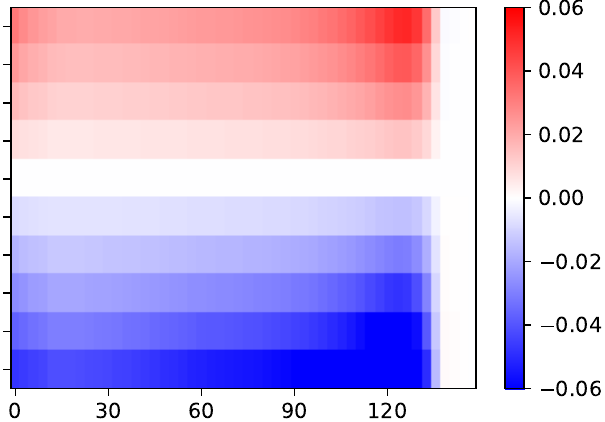} &
     \includegraphics[width = 0.22\textwidth, trim = 40 10 30 25, clip, page = 6]{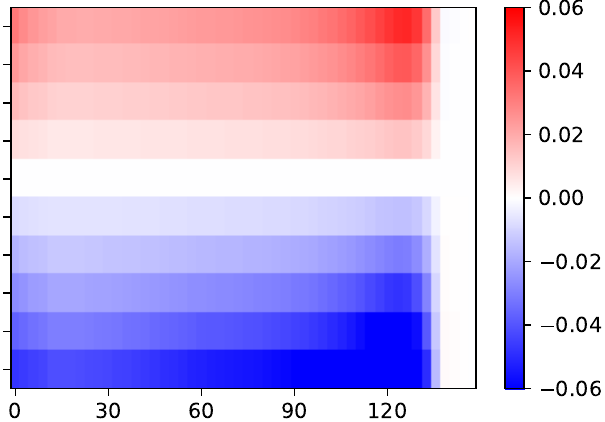} 
     &&\\
     \multicolumn{7}{c}{\footnotesize\textbf{Radius[pc]}}\\

    \end{tabular}
    \caption{Effects of the nuclear eject fraction on the [Fe/H], [Mg/Fe], [Eu/Fe] gas (over) abundance in the NSD. The  columns are the same as in Fig.~\ref{fig:ejectglobAbundance}}
    \label{fig:ejectnucAbundance}
\end{figure*}

Subtracting the median profile (in column 4 of Fig.~\ref{fig:ejectglobAbundance}) allows us to track the changes in oberabundance undistracted by the specific abundance pattern in each case and shows that the changes in overabundance are a simple consequence of the outer disc abundance change.
We see again that in the inflow scenario effects (e.g. here from the galactic disc ejection fraction) build up towards the inside and are overall retained longer, while they are drowned out by onfalling gas in the onflow case.
As expected from such a baseline effect, the changes in overabundance are almost an order of magnitude smaller than the changes in absolute abundance which stresses another time that the abundance difference to the bar tips is key to understanding NSD chemistry.

\subsubsection{NSD Ejection Fraction}

For comparison, we now look at the effect of the NSD ejection fraction (Fig.~\ref{fig:ejectnucAbundance}). The general effect is easy to understand: If less of the yields from dying stars are retained, the corresponding ISM abundances decrease. This is true for all the considered elements.
As this parameter only affects the NSD, taking the overabundance does not wash it out and its consequences are fully seen in the overabundances vs. the bar tips (2nd and 4th column).

The right-most column of Fig.~\ref{fig:ejectnucAbundance} shows that, to first order, lower NSD ejection fractions just result in correspondingly larger overabundances. To second order, in the inflow model, a larger NSD ejection fraction reduces the present gas mass as the inflow speed is limited in our model. However, from the figure we judge this effect on the abundances as rather minor.

\FloatBarrier 

\section{Conclusion}
\label{sec:Conclusion}

We have presented the first spatially resolved chemodynamical model for nuclear stellar discs, embedded in a full galactic chemical evolution model. The model incorporates an exponential nuclear gas disc, a nuclear ring with intense star formation on the edge of the NSD and a flow of gas funnelled down from the tips of the bar region, sustaining the intense star formation in the NSD.

Such chemical evolution models constrain the star formation history and genesis of a system. They translate the flows of gas, i.e. the accretion of gas, the loss of stellar yields, and internal (radial) flows into observable time- and radius-dependent abundance patterns, the simplest of which is the present-day abundance profiles of the cold star-forming gas.

As a first study, we have predicted and examined the main features expected for the gas abundance profiles in the NSD in different scenarios for the gas flow. We have dissected the enrichment history and found that the NSD chemical history is dominated by the lasting effect of the strongly star forming, outwards moving nuclear ring.  We have also shown that one should study the abundances of the NSD relative to the bar tips from where the gas is funnelled.

Our models show a global, mild enrichment of heavy element abundances relative to the tips of the bar from which gas is flowing in, and an enrichment particularly in $\hbox{[Mg/Fe]}$ and $\hbox{[Eu/Fe]}$ at the dense star-forming nuclear ring.\footnote{One caveat is needed here: If there is more gas from the bar/bulge, or if somehow significant amounts of gas can be accreted onto the NSD (a scenario we currently see no reason to adopt), this would result in overall lower metallicity (currently only the outermost fringe in our model NSD is slightly lower than the bar), but we do not expect major changes in the abundance ratios.} This peak is caused by the sudden rise of star formation efficiency caused by the infalling gas and the outward expansion of the star-forming ring. As $\alpha$-element enrichment is near-instantaneous, and iron enrichment from SNIa lags by ${\sim}1 \,{\rm Gyr}$, this ring structure is followed by a dip in $\hbox{[Mg/Fe]}\xspace$ at slightly smaller radii (where the star-formation rate is dropping). Europium, an r-process element from neutron star mergers, has an intermediate production timescale. Hence, the corresponding  peak-dip structure is slightly shifted to smaller radii, with the possibility of additionally lowered $\hbox{[Eu/Fe]}\xspace$ ratios on the outer edge of the ring, where ccSN contribute iron, before the NSM driven r-process yields kick in. 

Our main focus in this work is showing how two contrasting flow patterns in the NSD affect the results: an onfall-scenario, where the gas is funnelled from the bar onto all parts of the NSD directly, and an inflow scenario, where the funnelled gas is deposited on the nuclear ring and then flows inwards through the NSD. Those are to be understood as edge cases with the reality likely lying somewhere in between.
The scenario show marked differences. In particular, the radially inward flow of gas in the inflow model distributes the radial structures of abundance ratios further inwards and leads to a pile-up of heavy elements towards the inner NSD. This results in a significant gradient of NSD metallicity, which is much weaker or absent in the onfall model. 

We have also shown that the overall enrichment history and abundance profile of the galactic disc affects the NSD as the latter just reprocesses the funnelled-in gas. For example, the initial rise in galactic metallicity and decline in $\hbox{[Mg/Fe]}\xspace$ ($\hbox{[Eu/Mg]}\xspace$) ratios translates into lower NSD metallicity and higher $\hbox{[Mg/Fe]}\xspace$ ($\hbox{[Eu/Mg]}\xspace$) for an earlier NSD formation time. This shift directly translates to the measurements against the current bar tips.
As time progresses, the bar tips expanding further into the negative radial metallicity gradient implies a slow decrease in metallicity of the funnelled-in gas, which will be mirrored in the nuclear stellar disc by a negative stellar radial metallicity gradient.

Ultimately, understanding the NSD will involve major studies of parameter variations in models like ours. As an example, we have here studied how the galactic disc yield ejection fraction and the yield ejection fraction in the NSD affect its present-day abundances. This study particularly underlined the value of studying abundances of the NSD relative to the bar tips, which largely eliminate the impact of nuisance outer disc chemical evolution parameters like its ejection rate and helps study the chemical evolution of the NSD with better focus. We also note that such a relative abundance study will reduce the large systematic biases in abundance measurements.

In the future, models, like the presented, will serve to constrain galactic gas flows and structure. To just name two examples: we currently know little about the gas flows into the galactic centre and their long-term angular momentum balance, as only a fraction of the funnelled gas actually ends up in stars (and much less in the central black hole). A major part will be expelled in winds and we have shown that this kind of gas loss can be directly quantified by abundance differences. Secondly, the NSD is a unique laboratory with very little space for retaining hot gas close to the NSD, which implies that most of its hot gas should be expelled in the galactic fountain. This in turn means that by comparing nuclear stellar disc data with our models, we expect to be able to break the degeneracies in assessing what stellar types contribute in which fraction to the cold/hot gas phases, central to understanding early galactic enrichment. We will explore this topic further in a future paper.

Additionally, this paper has discussed the abundances of the star-forming gas and not of the stellar populations themselves. These require not only a summation in time (which can be done with RAMICES II), but also an additional treatment of radial mixing in the NSD. Comparisons with spectroscopic surveys thus have to be done with this in mind. More importantly, those surveys require careful modelling of the selection function, which is affected by the strongly varying dust extinction towards and within the NSD/CMZ. While measurements of the extinction value vary strongly \citep[e.g.][]{scoville_hubble_2003, schodel_peering_2010, nogueras-lara_galacticnucleus_2018}, they are  expected to be ${\gtrsim}\,3.35 \,{\rm mag}$ \citep{nogueras-lara_galacticnucleus_2021} which considerably shapes the observable stellar population. We will treat this in depth in our next paper.

Finally, there is plenty of caveats that should be named in such a work. Of course, the usual caveats of chemical evolution models, e.g. assumptions of stellar yields, IMF assumptions, etc. should be regularly revisited. Those are important, but we expect them to have lesser impact on the statements we make in this paper. The more unsettling parts are the uncertainties in both theory and observations: On the theoretical side, little modelling has been done for NSDs, especially of the stellar contribution. In particular, it is unclear how stars in the NSD get re-distributed e.g. in radius: there is some spiral structures observed, but there is currently no measure for radial migration, heating, and related processes. Due to this uncertainty we here showed mainly abundance profiles for the star-forming gas for which stellar migration plays a subordinate role compared to gas flows (as shown in many galactic disc papers).

Additionally, it remains unknown how much gas from the bar area (e.g. stellar yields, coronal accretion) is swept up during the funnelling and pollutes the inflowing gas.

Further, we make our predictions depending on parameters currently in a rather observation-free space, as we lack resolved, high-quality observations for both stellar abundance patterns and also present-day star-forming gas abundances in the NSD. So, this paper should also be taken as a strong encouragement to push for better data: NSDs can help in constraining important galactic processes, but this is only possible once we can challenge models like ours with high-quality data.

\begin{acknowledgements}
   We thank Jack Fraser-Govil, Mattia Sormani and Mike Rich for helpful discussions and Mathias Schultheis for providing us with the metallicity distribution function of the KMOS NSD survey. 
   We also thank the anonymous referee at A$\&$A for helpful comments. 
   JF acknowledges support from University College London’s Graduate Research Scholarship. This work has received funding from the European Research Council (ERC) under the Horizon Europe research and innovation programme (Acronym: FLOWS, Grant number: 101096087).
   RS thanks the Royal Society for generous support via a University Research Fellowship.    
\end{acknowledgements}

\bibliographystyle{aa}
\bibliography{references}

%
%
\begin{appendix} 
   \section{Summary of the Chemical Evolution Prescription}
\label{sec:appChemEv}
\begin{figure}
    \centering
    \includegraphics[width = 0.9\columnwidth]{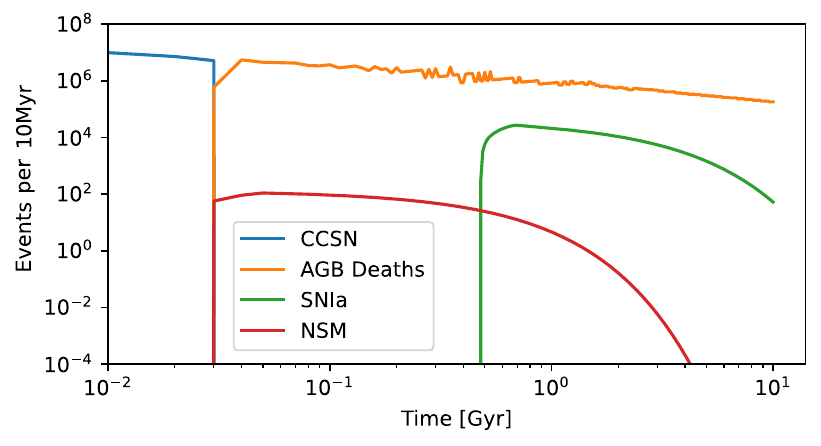}
    \caption{Event rates of different stellar death scenario over time for a RAMICES II (see Sec.~\ref{sec:Model}) model without star formation except for the very first timestep. The temporal resolution is 0.01 Gyr, i.e. ccSN only play a role for the first 3 timesteps in this example. The simulation contains $10^{11}\,{\rm M}_\odot$ of primordial gas.}
    \label{fig:Eventrates}
\end{figure}

\begin{sidewaystable*}
    \caption{Adopted yield grids and prescriptions for the timescales important to chemical evolution modelling.}
    \centering
        \bgroup
        \def\arraystretch{1.5}
        \begin{tabular}{c||c|l|c}
        Process & Shape & Values & Yields \\
        \hhline{=||=|=|=}
        \multirow{4}{*}{Star formation} & \multirow{4}{*}{$\dot{\Sigma}_\mathrm{*} =\begin{cases}\kappa\Sigma_\mathrm{gas}^{n1} & \Sigma_\mathrm{cold} > \Sigma_\mathrm{cut-off}\\\kappa(\Sigma_\mathrm{cut-off} + \Sigma_\mathrm{hot})^{n_1-n_2} \Sigma_\mathrm{gas}^{n_2} & \text{otherwise}\end{cases}$} & $n_1 = 1.4$ &\\
         &  & $n_2 = 4.0$ &\\
         &  & $\kappa = 2.2 (10^9\,{\rm M}_\odot)^{-n_1+n_2} \,{\rm Gyr}^{-1}$ &\\
         & 
         & $\Sigma_\mathrm{cut-off} = 0.004 \cdot 10^9\,{\rm M}_\odot \,{\rm kpc}^{-2}$&\\
         \hline
        ccSN $\&$ AGB & \makecell{PARSEC isochrones (1,2,3)  \\including the thermally pulsing-AGB phase (4,5,6)}& $\tau_\mathrm{PARSEC}(M_*,Z)$ &\makecell{(7,8,9) }\\
        \hline
        \multirow{4}{*}{SNIa} & \multirow{4}{*}{
            \makecell{$\Psi_\mathrm{SNIa}(t) = 
            \begin{cases}
                0 & t< t_\mathrm{SNIa, delay}\\
                f_\mathrm{SNIa, short} \mathrm{e}^{-t/\tau_\mathrm{SNIa,short}} + (1-f_\mathrm{SNIa, short})\mathrm{e}^{-t/\tau_\mathrm{SNIa, long}}&\text{otherwise} 
            \end{cases}$} }
                & $t_\mathrm{SNIa, delay} = 0.45 \,{\rm Gyr}$ &\multirow{4}{*}{\makecell{W70 yields (10)}}\\
         &  & $f_\mathrm{SNIa, short} = 0.01$  & \\
         &  & $\tau_\mathrm{SNIa, short} = 100 \,{\rm Myr}$& \\
         &  & $\tau_\mathrm{SNIa, long} = 1.5 \,{\rm Gyr}$ &\\
         \hline
        \makecell{NSM} & \makecell{
            $\Psi_\mathrm{NSM}(t) =
            \begin{cases}
                0 & t< t_\mathrm{NSM, delay}\\
            e^{-t/\tau_\mathrm{NSM}}& \text{otherwise}                 
            \end{cases}$} & \makecell[l]{ $t_\mathrm{NSM, delay} = 20 \,{\rm Myr}$ \\\vspace{-6pt}  \\ $\tau_\mathrm{NSM} = 300 \,{\rm Myr}$} &\makecell[c{p{3.4cm}}]{unclear, a fixed Eu yield is assumed and calibrated to the solar neighbourhood abundance by scaling the fraction of neutron stars undergoing a  merger at each time step}\\
         
         \hline
        cooling & $\dot{M}_\mathrm{h, cooling} = - \Lambda M_\mathrm{h}$ & $\Lambda = 1 \,{\rm Gyr}$ &\\
        \hline
        \multirow{5}{*}{\makecell{inside-out \\ formation}} & \multirow{5}{*}{$R_\mathrm{gas}(t) = R_0 + N \left[\arctan\left(\frac{t-t_0}{t_g}\right) -\arctan\left(\frac{t_0}{t_g}\right) \right]$} & $R_0 = 1.5 \,{\rm kpc}$ &\\
         &  & $R_\mathrm{end} = 3.75 \,{\rm kpc}$ &\\
         &  & $t_0 = 1.0 \,{\rm Gyr}$ &\\
         &  & $t_g = 0.6 \,{\rm Gyr}$ &\\
         &  & $t_\mathrm{end} = 12 \,{\rm kpc}$ &\\
        \end{tabular}

        \egroup
            
    \tablebib{(1) \cite{bressan_parsec_2012}, (2) \cite{chen_parsec_2015}, (3) \cite{tang_new_2014}, (4) \cite{marigo_new_2017}, (5) \cite{pastorelli_constraining_2019}, (6) \cite{pastorelli_constraining_2020}, (7) \cite{maeder_stellar_1992}, (8) \cite{marigo_chemical_2001}, (9) \cite{chieffi_explosive_2004}, (10) \cite{iwamoto_nucleosynthesis_1999}.}
    \label{tab:TimeScaleValues}
\end{sidewaystable*}

Here we summarise the main chemical evolution timescales and formalise them in the way they are implemented in RAMICES II \citep[see][]{fraser-govil_advancements_2022}.
Table \ref{tab:TimeScaleValues} summarises the main prescriptions. Fig.~\ref{fig:Eventrates} illustrates how the different timescales translate to event rates for each single stellar population. First a fraction of the heavy stars with short lifetimes explode as ccSN. As this is only the case for stars with $M \gtrsim 8\,{\rm M}_\odot$ which have a lifetime of $\lesssim 0.1\,{\rm Gyr}$, the ccSN rate quickly peaks and then ceases.

A small fraction of the resulting neutron stars in binary systems spiral in and eventually merge. RAMICES II uses a delay time of 20 Myr followed by decay distribution of $e^{-t/300\textrm{ Myr}}$ after the formation of the neutron stars.  

After the ccSN cease, intermediate/lower mass stars start to reach the end of their lives and eject their envelopes. Even though the IMF favours low-mass stars, their longer lifetime leads to a declining rate of AGB-star deaths and yields.

Lastly, SNIa require existing white dwarfs. We employ an initial delay time of $0.45\,{\rm Gyr}$ after WD formation and model SNIa rates with a two-exponential delay distribution composed of a fast term ($e^{-t/100\,{\rm Myr}}$) for 1$\%$ of the SNIa and standard $e^{-t/1.5\,{\rm Gyr}}$ for the rest.

While our models do not include a comprehensive thermodynamical treatment of the gas, important physics would be lost without considering cooling in the ISM. RAMICES II hence implements a simplified two-phase model with distinct hot and cold gas, and a constant cooling timescale $\Lambda = 1 \,{\rm Gyr}$: $\dot{M}_\mathrm{cool} = - \Lambda M_\mathrm{h}$, with $M_\mathrm{h}$ being the hot gas mass.

RAMICES II adopts a fully differentiable function to trace inside-out formation following \citet{schonrich_understanding_2017}.

\begin{figure}
    \centering
    \includegraphics[width = \columnwidth]{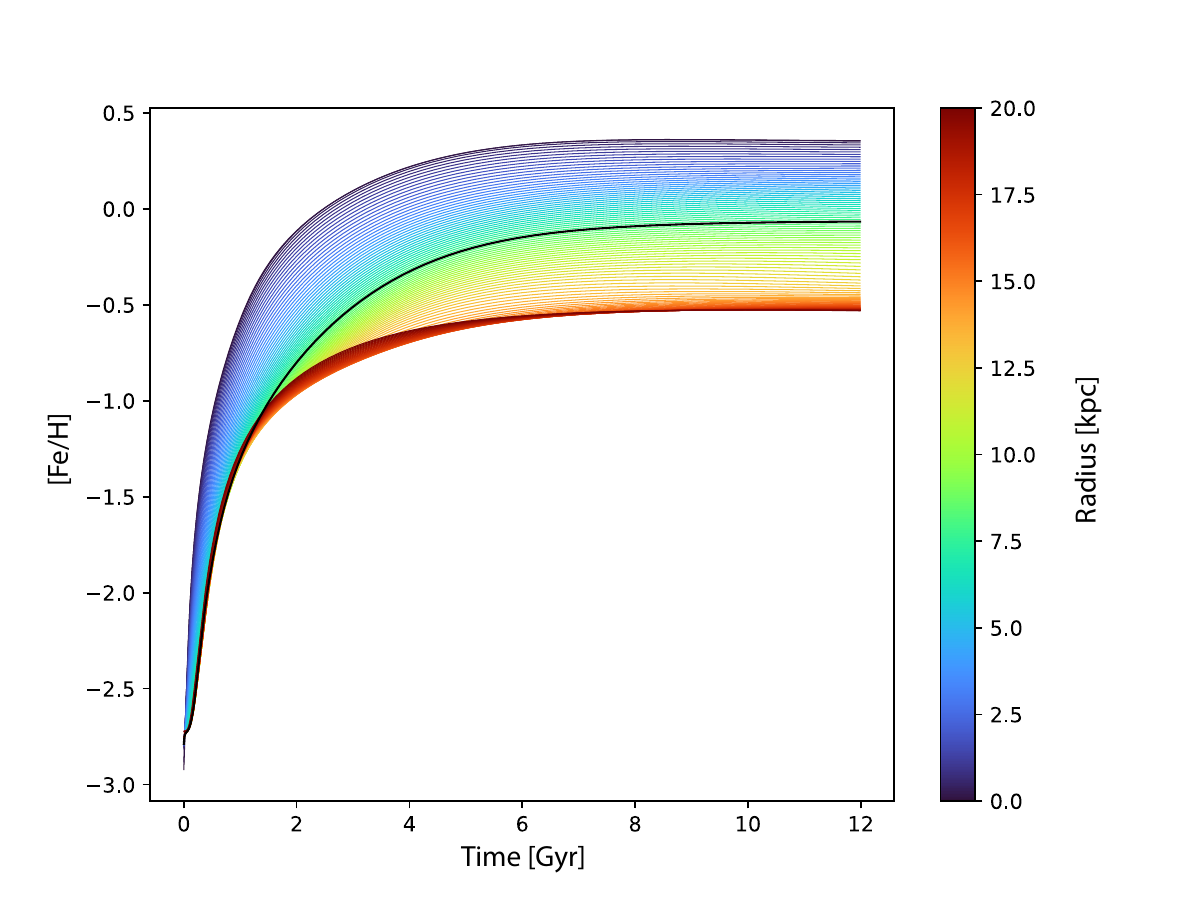}
    \caption{Global iron abundance over time for our best fitting RAMICES II model. The abundance of the solar annulus is highlighted in black. }
    \label{fig:GlobalIronAbundance}
\end{figure}

The resulting evolution of the metallicity in the galactic model as a function of radius (colour coded) and time (x-axis) is shown in Fig.~\ref{fig:GlobalIronAbundance}. Two main features matter for this paper: $\hbox{[Fe/H]}\xspace$ shows an initial rise towards a plateau covering the last $\sim 8$ Gyrs of evolution. The initial rise may be inherited by the NSD if bar formation falls into that period. Further, the strong radial abundance gradient of order $-0.06 \,{\rm dex} \,{\rm kpc}^{-1}$ in concordance with Cepheid data \citep{luck_distribution_2011} imprints on the bar/NSD: as the bar grows, the bar tips move outwards, reducing the metallicity of the material feeding inwards and - if the NSD grows inside-out - it can thus inherit a negative radial metallicity gradient.

\section{Data situation}
\label{sec:dataSituation}
\begin{figure}
    \centering
    \includegraphics[width = \columnwidth]{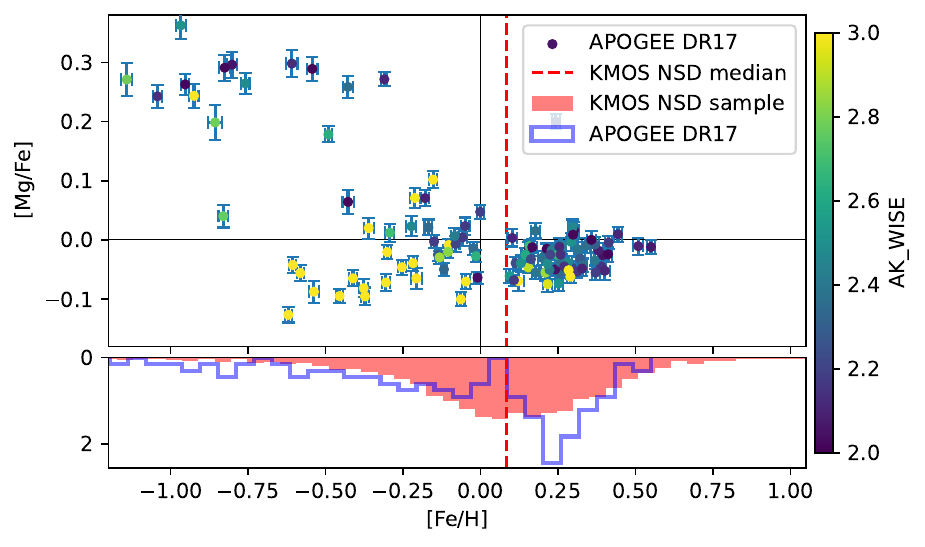}
    \caption{Comparison of the APOGEE DR17 \citep{abdurrouf_seventeenth_2022} stars with WISE reddening AK$\_$WISE $> 2$ and passing recommended quality checks (see text), and the KMOS NSD sample from \citet{fritz_kmos_2021}.}
    \label{fig:ApogeeKmos}
\end{figure}

Fig.~\ref{fig:ApogeeKmos} shows the problematic situation with current Milky Way NSD data from APOGEE vs. KMOS \citep{abdurrouf_seventeenth_2022, fritz_kmos_2021}. From the APOGEE sample we chose stars within $\Delta l = 1.3 ^\circ$ and $\Delta b = 0.4 ^\circ$ of the Galactic centre, excluding stars with  ASPCAPFLAG==M$\_$H$\_$BAD or STAR$\_$BAD and stars with flags 2, 3, 9 in STARFLAG set (VERY$\_$BRIGHT$\_$NEIGHBOR, LOW$\_$SNR, PERSIST$\_$HIGH). A full description of the APOGEE flags can be found in \citet{holtzman_abundances_2015}. This leaves 132 stars, 113 of which also include Mg abundances. It is obvious that the selected APOGEE stars' metallicity distribution does not match that from KMOS. We also found that APOGEE shows a strong negative correlation for the entire disc between $\hbox{[Mg/Fe]}\xspace$ and metallicity vs. extinction, suggesting analysis problems. This is not surprising, as stars with high metallicity, low temperature and high extinction are particularly difficult to analyse. The bottom line is that APOGEE abundances should currently not be used for NSD stars and in high extinction regions in general.

\section{Effects of other chemical evolution parameters}
\label{sec:appendixExtraMaterial}
Here we examine some chemical evolution parameters that we deem of lesser importance, but whose effects the reader might wonder about.

\begin{figure*}
    \centering
    \includegraphics[width = 0.9\textwidth]{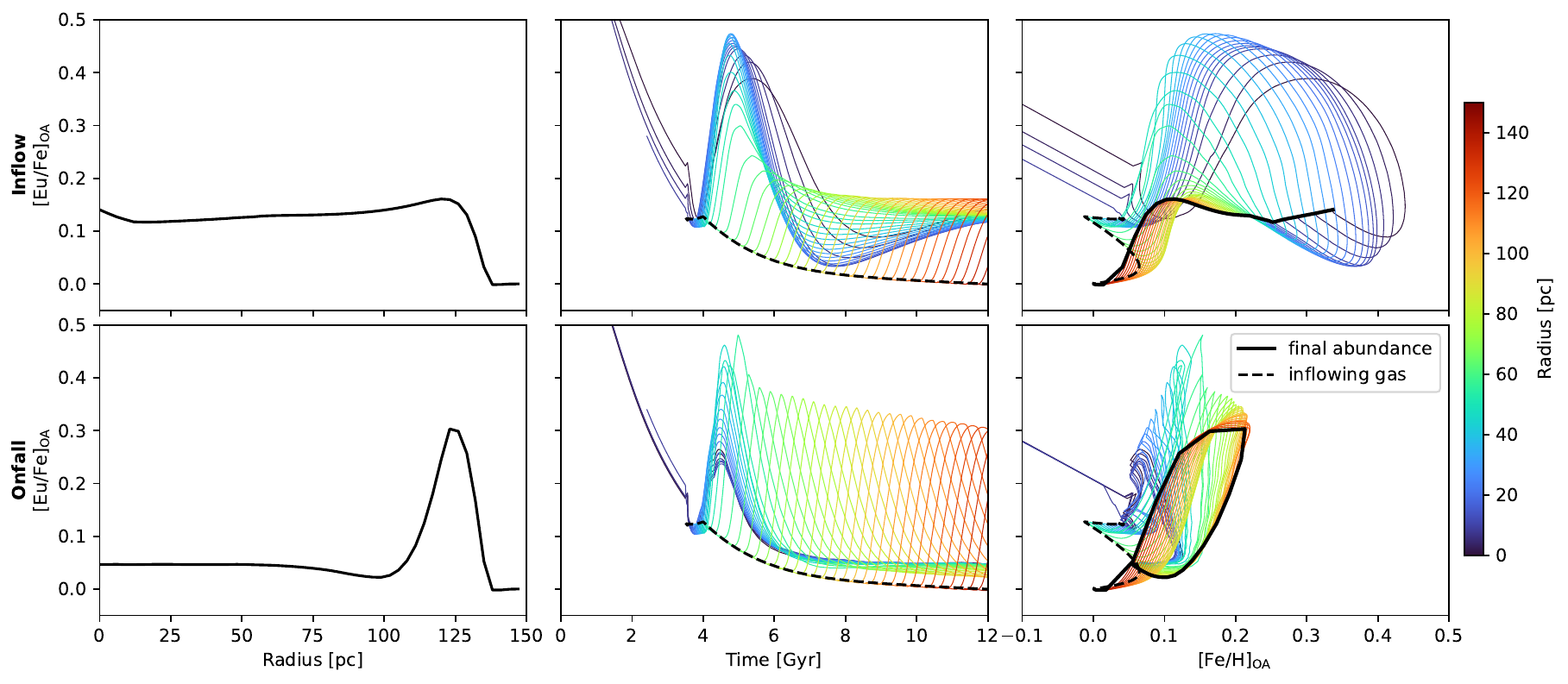}
    \caption{Comparison of the inflow and onfall scenario in [Eu/Fe]$_\mathrm{OA}$}
    \label{fig:metplaeEufe}
\end{figure*}

\subsection{[Eu/Fe] abundances} 

Fig.~\ref{fig:metplaeEufe} adds to the NSD detailed abundance models shown in Figs.~\ref{fig:gradient} and \ref{fig:gradientInflow} by showing in addition the $\hbox{[Eu/Fe]}\xspace$ ratios (instead of $\hbox{[Eu/Mg]}\xspace$) for both the Onfall (top row) and Inflow (bottom row) scenarios. There is no substantive difference to the plots discussed in the main text, except for a slight shift of the features from the longer NSM onset time.

\subsection{Surface Density}
\begin{figure*}
    \centering
    \includegraphics[width=1.0\textwidth]{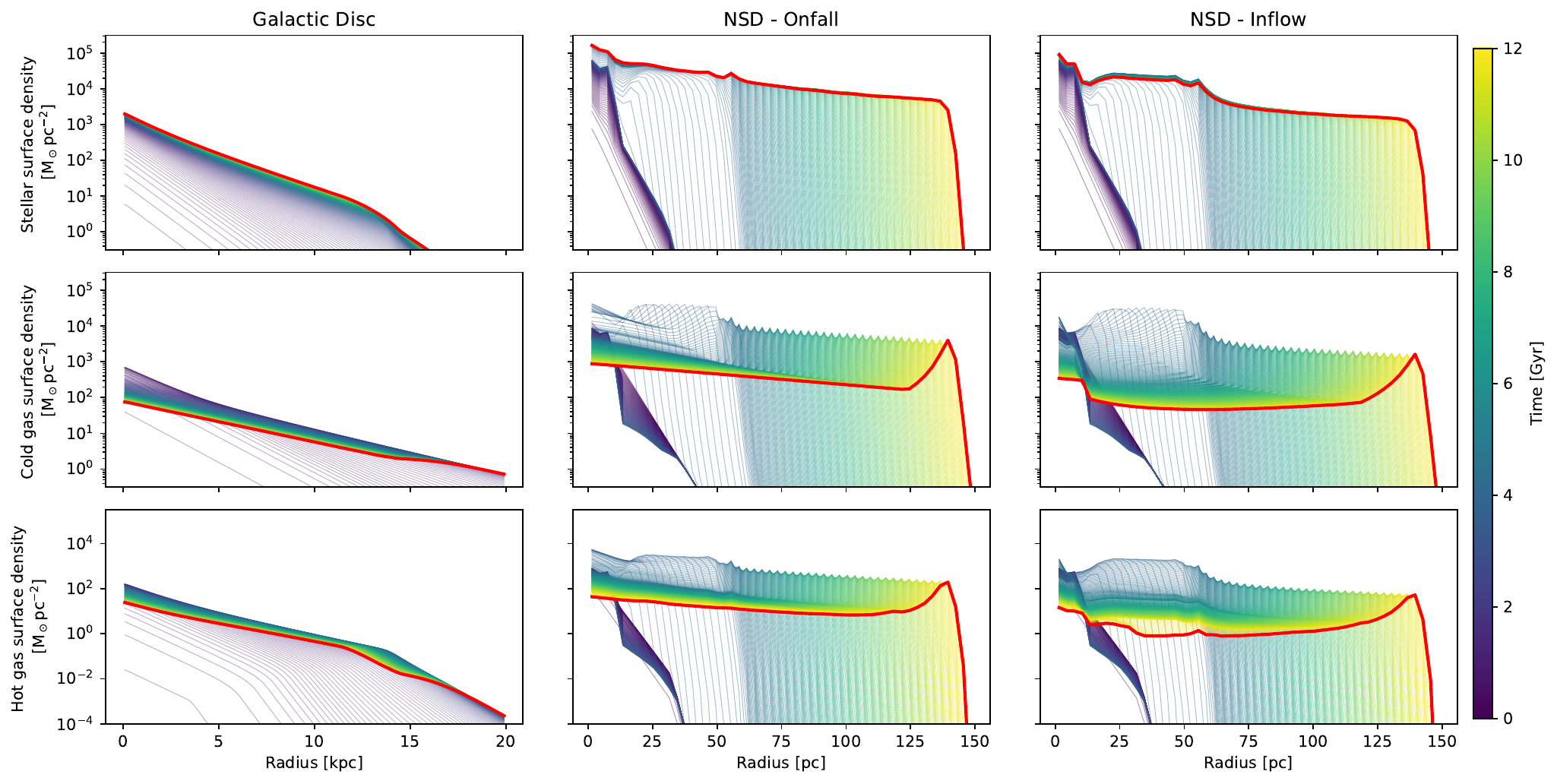}
    \caption{Surface density profiles of the stars, hot and cold gas mass in our model, using the same data as Fig.~\ref{fig:gasmass}. The left column shows the surface densities for the full galactic disc, the two NSD accretion scenarios are on the right and the thick red line shows the density profile at the end of the simulation.}
    \label{fig:gasmassSurfacedensity}
\end{figure*}

Fig.~\ref{fig:gasmassSurfacedensity} shows the same data as the upper three rows of Fig.~\ref{fig:gasmass}, but in terms of surface density. The NSD appears as a very dense object (both in stars in the  upper row as well as in the gas in the bottom two rows) whose stellar component is approximately exponential. We also see the difference between the inflow and onfall scenarios (the two right columns): Due to our limit on gas inflow velocity (see also Sec.~\ref{sec:maxsteal} below), the  NSD gaseous disc in the inflow scenario cannot always reach the shape prescribed by eq.~\eqref{eq:surface_density} and is hence not completely exponential.

\subsection{Mixing strength}
\begin{figure*}
    \centering
    \includegraphics[width=1\textwidth]{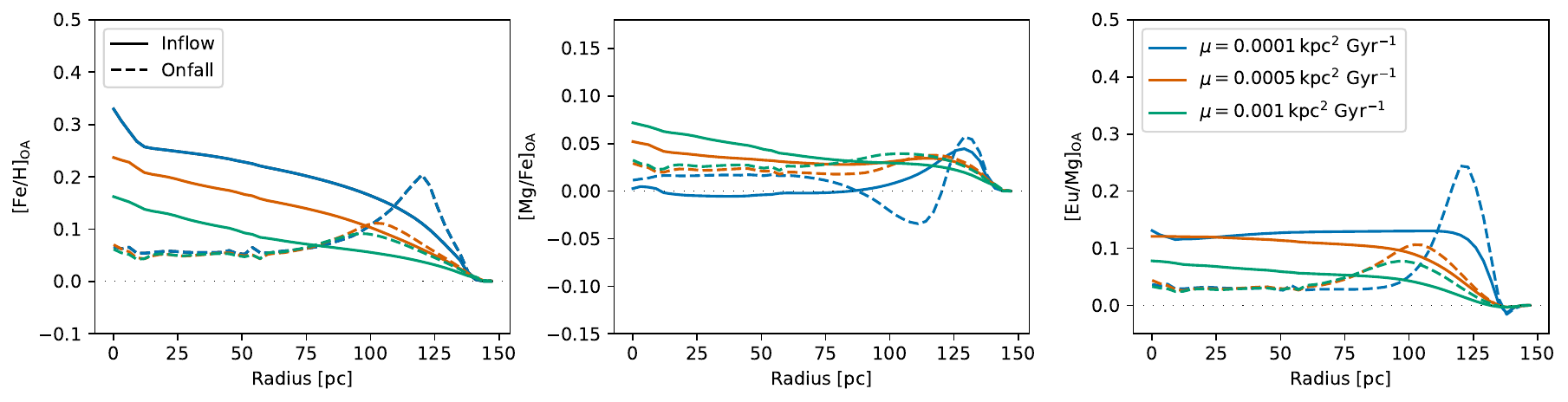}
    \caption{Final NSD overabundances for different values of the mixing parameter.}
    \label{fig:MixingStrength}
\end{figure*}

So far in this work, we did not take into account radial migration of stars within the NSD. The NSD was specifically identified as a dynamically cool component \citep[e.g.][]{launhardt_nuclear_2002,  schonrich_kinematic_2015, sormani_stellar_2022}, kinematically distinct from the NSC in the centre as well as from the surrounding bulge. \citet{schultheis_nuclear_2021} corroborated this by showing that the metal rich stellar component in the NSD area (chemically distinct from the surrounding bulge) also has a significantly lower velocity dispersion than the metal poor stars. 
Therefore, the majority of NSD stars are expected to be on nearly circular orbits. However, this might trigger instabilities like an inner bar or spiral arms, driving radial migration. 
The currently available data does not allow us to draw conclusions on the prevalence or magnitude of this effect for the Milky Way NSD. 

Hence, we study here the effects of parametrised radial migration on the NSD abundance profiles, giving three models of varying mixing strength in \autoref{fig:MixingStrength}. RAMICES II models stellar radial migration through a discrete time Markov chain, i.e. with a migration probability at each timestep independent of its previous migration history: $\propto \frac{\mu}{\Delta R^2} $, where $\mu$ is the diffusion parameter and $\Delta R$ is the model ring width \citep[cf. chap. 4.5][]{fraser-govil_advancements_2022}. 

\autoref{fig:MixingStrength} shows the final gas overabundances for three models with diffusion parameters $\mu =  0.0001 \text{ kpc}^2\text{Gyr}^{-1}$, $0.0005 \text{ kpc}^2\text{Gyr}^{-1}$ and $0.001 \text{ kpc}^2\text{Gyr}^{-1}$. This corresponds to radial migration mixing times $t_{mix} \approx R^2/\mu$ of 225 Gyr, 45 Gyr and 22.5 Gyr (corresponding to the blue, green and orange lines; for comparison: the mixing time of the full galactic model is $\approx 1,000 \mathrm{Gyr}$). 
Experimenting with the parameters showed that the final overabundances do not significantly change for larger mixing times than $\approx 225$ Gyr (blue line in \autoref{fig:MixingStrength}; the fiducial model in this work uses $\mu = 10^{-9} \text{ kpc}^2/\text{Gyr}$ which gives a nearly indiscernible final abundance profile). 

In iron, shorter mixing times lead both to a shallower metallicity gradient (see the $\hbox{[Fe/H]$_\mathrm{OA}$}\xspace$-gradient in the inflow scenario) as well as a dampening of the effect of the nuclear ring as the formed stars get redistributed more quickly (seen well in the onfall scenario where the centre of the $\hbox{[Fe/H]$_\mathrm{OA}$}\xspace$-peak shifts to smaller radii). 
In contrast, in $\hbox{[Mg/Fe]$_\mathrm{OA}$}\xspace$ and $\hbox{[Eu/Mg]$_\mathrm{OA}$}\xspace$, the quicker redistribution of young, high-$\alpha$ stars from the nuclear ring even leads to a steepening of the abundance gradient.

In general, the effect of stronger mixing is most pronounced for the infall scenario - as the centre of the onfall models are dominated by gas directly accreted from the bar, the region $\lesssim 60$ pc is unaffected by the radial mixing.

In conclusion, while the radial profile in the gas (over)abundances gets dampened by a lower mixing time, the general shape of the abundance profiles is conserved. Especially in the onfall scenario, the shift of the abundance peak towards the GC is in principle testable and the difference between maximal $\hbox{[Mg/Fe]$_\mathrm{OA}$}\xspace$-abundance and peak in star formation might be a good measurement for the expected radial migration.

\subsection{Limiting Maximal Inflow Velocity}
\label{sec:maxsteal}
\begin{figure*}
    \centering
    \includegraphics[width=1.0\textwidth]{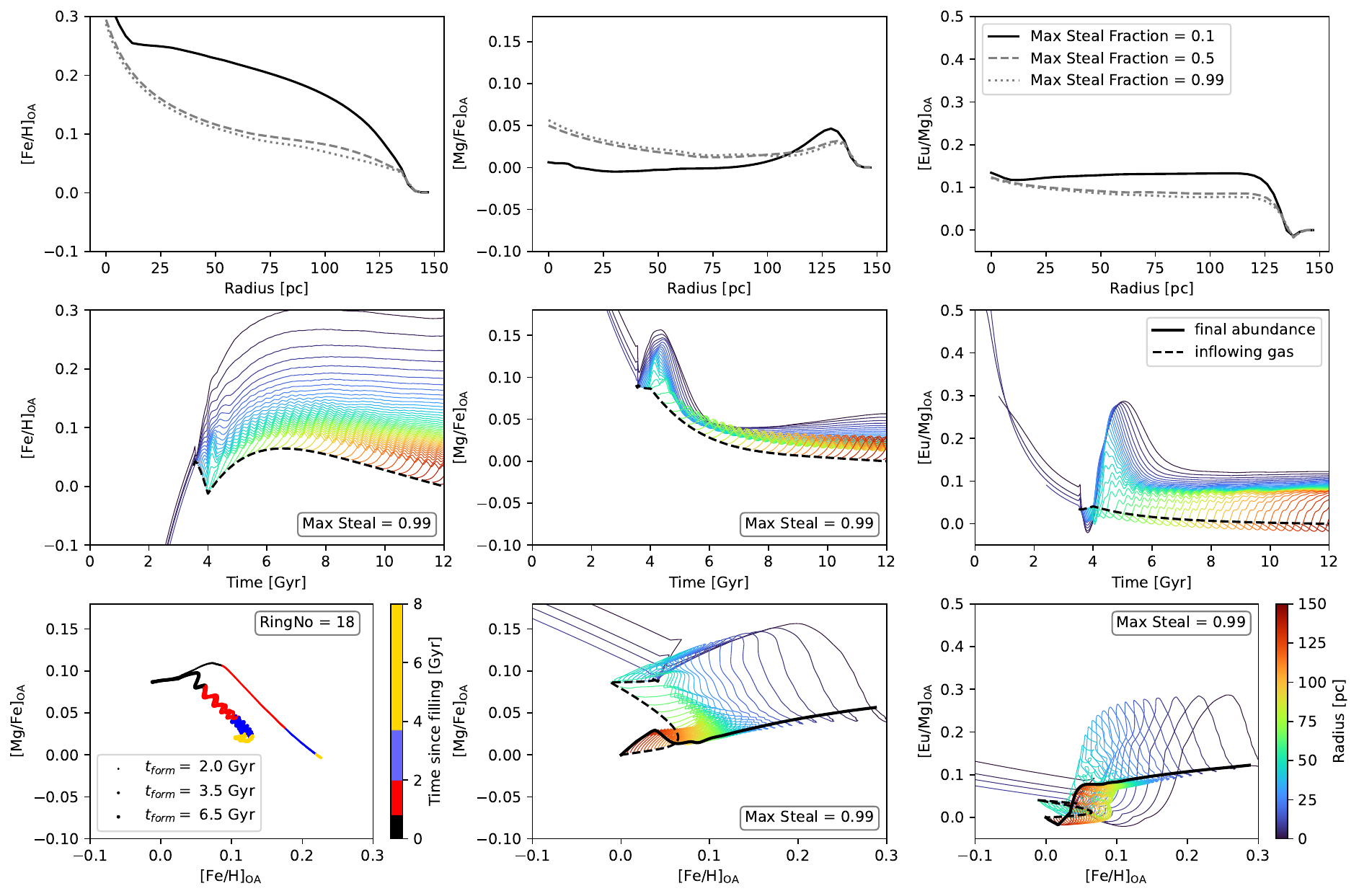}
    \caption{Following Fig.~\ref{fig:gradient}, we show the influence of a higher maximal inflow fraction on the NSD gas chemical evolution. The middle row and bottom middle and right panel hereby focus on a max steal fraction of 0.99 (i.e. a maximal inflow velocity of $\sim 0.3\,{\rm km}\,{\rm s}^{-1}$). The bottom left overlays the evolution of a single ring (at $R=55.5 \,{\rm pc}$) for all three treated max steal fractions. }
    \label{fig:metplanemaxsteal}
\end{figure*}

\begin{figure*}
    \centering
    \includegraphics[width=0.95\linewidth]{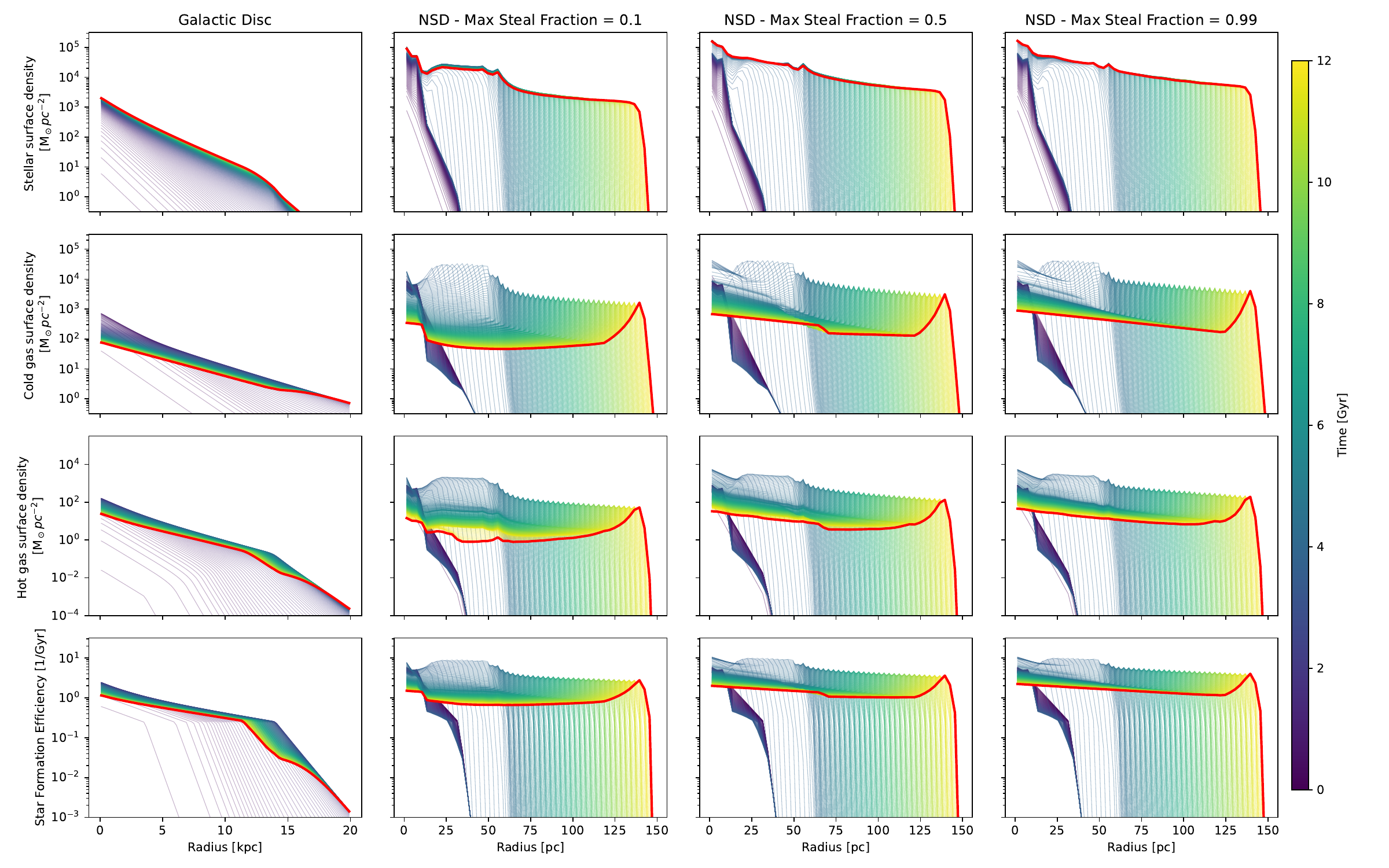}
    \caption{ Surface density of stars, cold and hot gas and star formation efficiency for the total galactic disc (left column) and the evolution of the NSD surface density for 3 different maximal inflow fractions (see the paper text for details). The thick red line in every panel shows the final surface density/star formation efficiency profile.}
    \label{fig:gasmassmaxsteal}
\end{figure*}

We do not directly prescribe radial flow velocities in our inflow model. However, the time and spatial resolution introduces a numerical/resolution limit: 
\begin{equation}
    v_\mathrm{i, max} = \frac{\Delta R}{\Delta t} = \frac{3 \,{\rm pc}}{10 \,{\rm Myr}} \sim 0.3 \,{\rm km}\,{\rm s}^{-1}.
\end{equation} 

In this case all gas in a ring flows to the next in each timestep. Such a strong flow can also lead to metallicity oscillations in the simulation.
Hence, we implemented a "maximal steal fraction" limiting the transfer of gas per timestep, effectively limiting the inflow velocity to $v_\mathrm{i, max} = f_{\rm max, steal} \frac{\Delta R}{\Delta t}$.

Without any restrictions imposed, this flow fraction is between $< 0.1$ (for the inner $25 \,{\rm pc}$) an  $\sim 0.6$ (just inside the nuclear ring). Hence, for any $f_{\rm max\, steal} \gtrsim 0.5$, the resulting abundances are indistinguishable, as shown in Fig.~\ref{fig:metplanemaxsteal}. 

Limiting the inflow velocity also limits the accreted gas mass at the nuclear ring and the total mass of the NSD (see Fig.~\ref{fig:gasmassmaxsteal}).

Comparing the shape of the abundance profiles affected by a larger gas inflow velocity to the effect of stronger mixing (Fig.~\ref{fig:MixingStrength}) shows that the former keeps the position of the abundance peak constant (the high-abundance gas gets only moved after it has been produced in stars) while the latter shows a smearing out of the abundance peak with the overall maximum also shifting (as the stars that produce the high-abundance gas themselves move before dying). 

As explained in Sec.~\ref{sec:Model}, the literature provides remarkably little constraints on NSD flows. 

On the theory side, recent MHD hydro-dynamical simulation \citep{tress_magnetic_2024} find a NSD surface density of $\gtrsim 100 \,{\rm M}_\odot/\,{\rm pc}^2$ after initial formation (their Fig.~6), and an inflow rate of $\dot{M} \sim 0.02 \,{\rm M}_\odot/{\rm yr}$ (their Fig.~22). With a radius of $\sim 100 \,{\rm pc}$, this leads to a velocity of $v_\mathrm{i} \simeq \dot{M}/(\Sigma R 2 \pi) \lesssim 0.3 \,{\rm km}\,{\rm s}^{-1}$ . However, those simulations might overestimate the inflow, as they for example do not include general outflow. 

Hence, for this work, we chose to limit $f_{\rm max\,steal}$ to $0.1$ in order to obtain a constant inflow velocity and median mixture, leading to a maximal inflow velocity of  $v_\mathrm{i, max} = 0.1\frac{\Delta R}{\Delta t} \sim 0.03 \,{\rm km}\,{\rm s}^{-1}$ and leave a further exploration of the inflow history to future work. 

\subsection{Bar Formation Epoch}
\label{sec:barformationepoch}
\begin{figure*}
    \centering
    \includegraphics[width=0.95\textwidth]{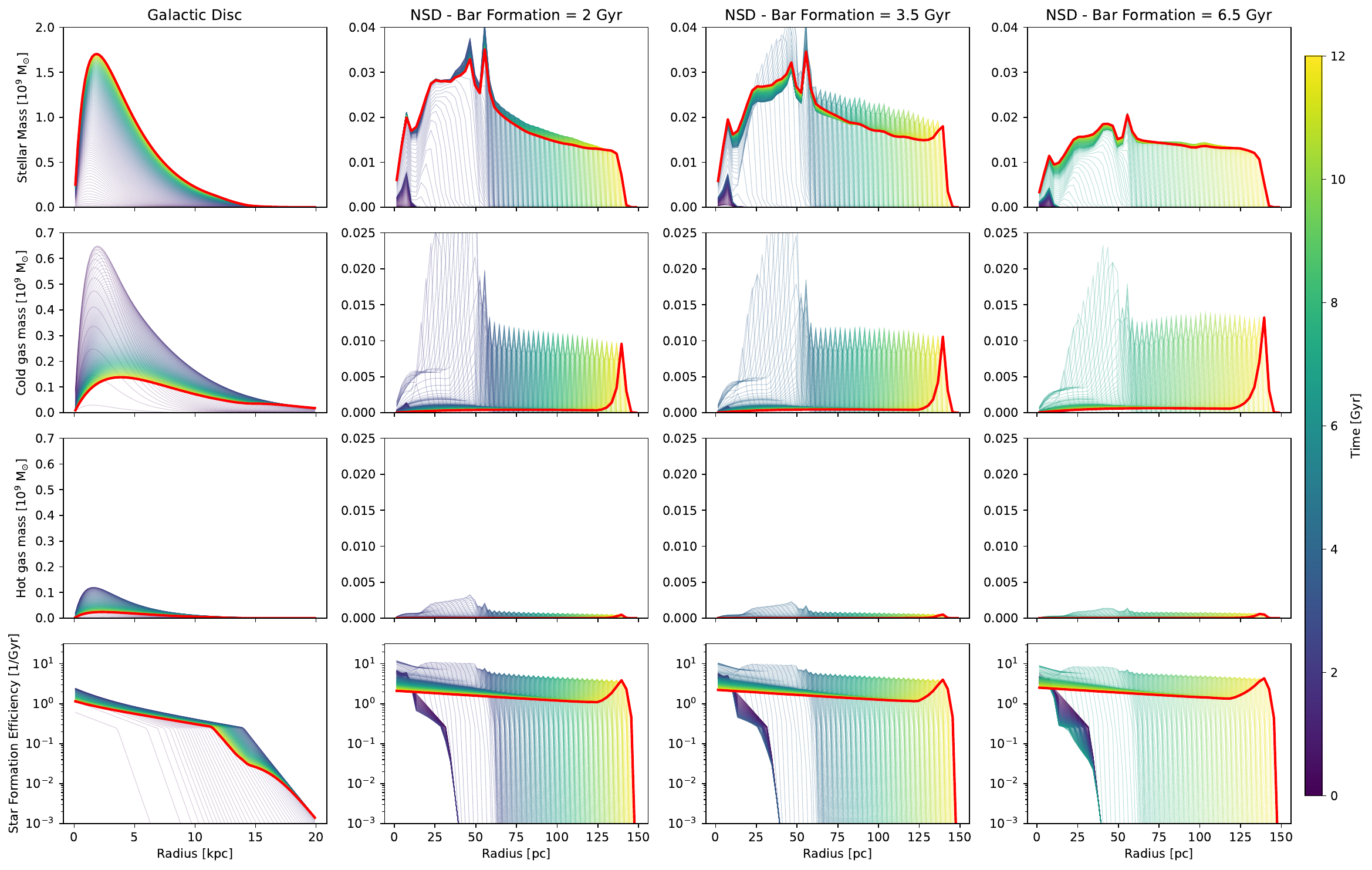}
    \caption{Following Fig.~\ref{fig:gasmass}. The right 3 columns show the evolution of the NSD gas mass for 3 different bar formation times in the onfall scenario.}
    \label{fig:gasmassbarformation}
\end{figure*}

\begin{figure*}
\centering
    \textbf{\large Onfall}\\
    \includegraphics[width=1\textwidth]{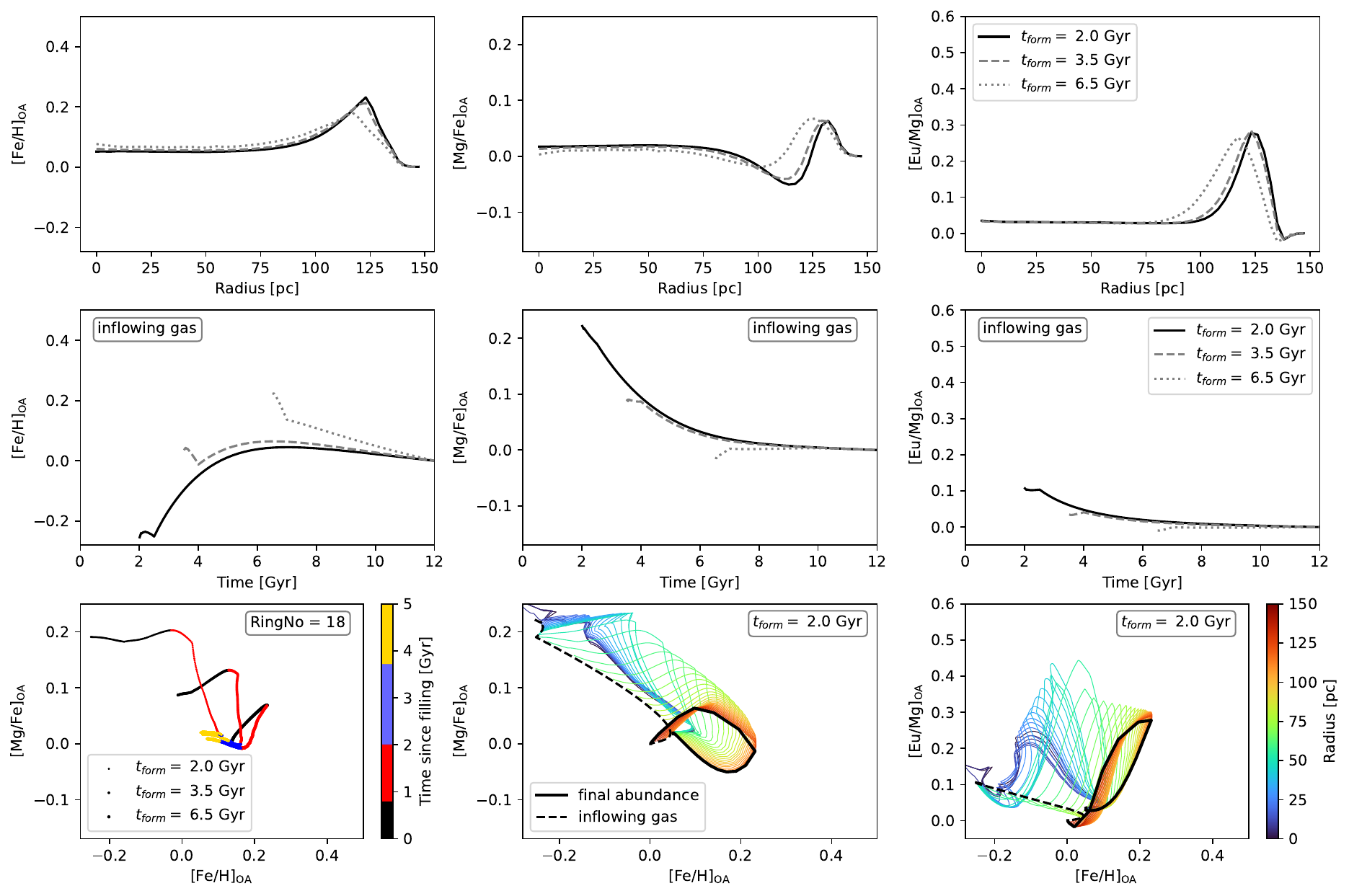} \\
    \hrule
    \vspace{6pt}
    \textbf{\large Inflow}\\
    \includegraphics[width=1\textwidth]{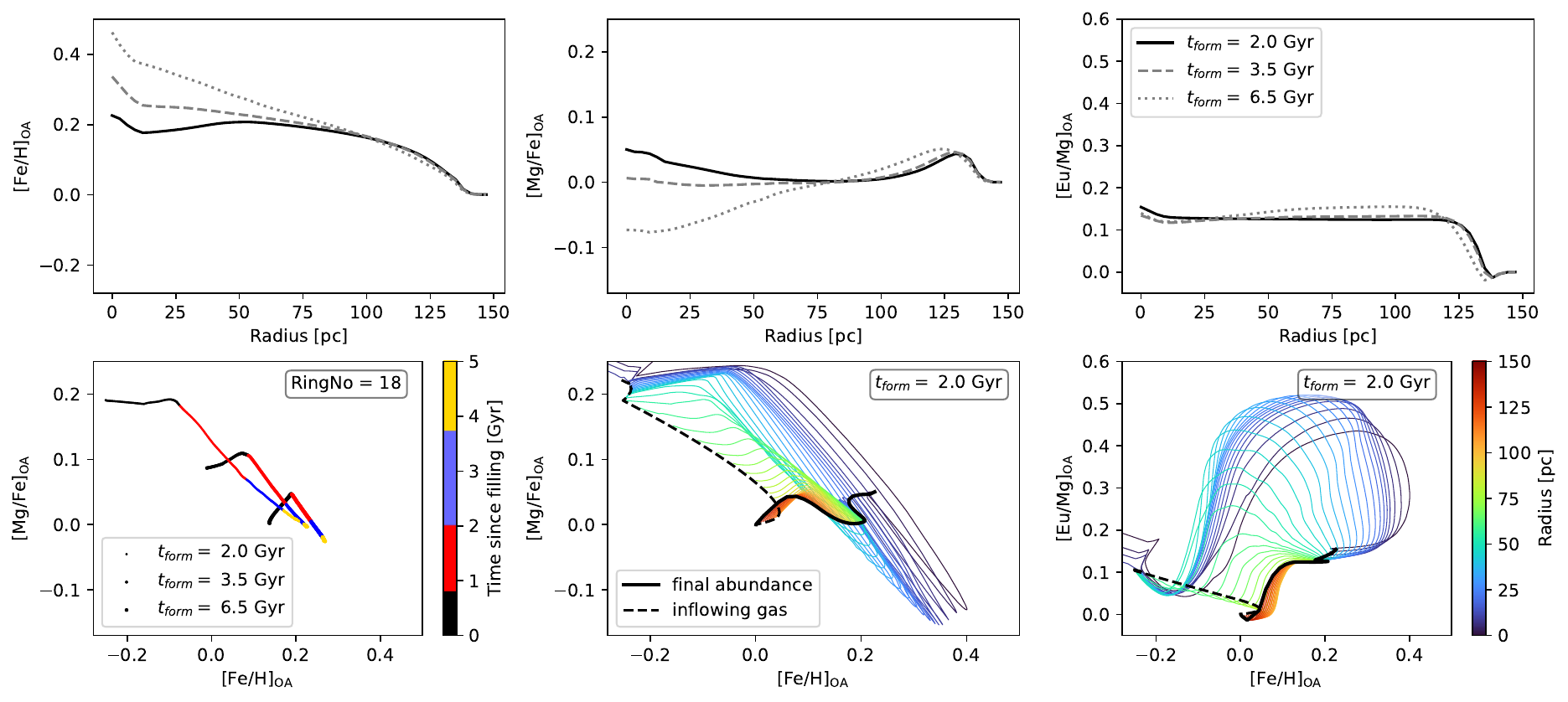}

    \caption{Following Fig.~\ref{fig:gradient} and \ref{fig:gradientInflow}, we show the influence of an earlier/later bar formation epoch on the NSD gas chemical evolution. The middle row of the onfall scenario shows the overabundance of the gas feeding the NSD and is the same for both models. 
    The  bottom middle and right panel hereby focus on an early bar formation time. The bottom left overlays the evolution of a single ring (at $R=55.5 \,{\rm pc}$) for all three considered formation epochs.}
    \label{fig:metplanebarformation}
\end{figure*}

The formation time of the Galactic bar is still under debate. \citet{baba_age_2020} proposed to use the oldest NSD stars, as the NSD formation epoch is expected to coincide with that of the Galactic bar. Using Mira variables, \citet{sanders_epoch_2024} timed the NSD formation time to $\tau \gtrsim 8\,{\rm Gyr}$. 

Therefore, we consistently choose a bar formation time $8.5 \,{\rm Gyr}$ ago. However, the results from \citet{sanders_epoch_2024} are also weakly consistent with an even earlier formation time. We ran two more models (with a bar formation time after $2 $ and $6.5 \,{\rm Gyr}$, i.e. $10 \mathrm{Gyr}$ and $5.5 \mathrm{Gyr}$ ago) to test the influence on NSD gas abundances, while keeping the initial and final size of bar and NSD fixed. Hence, in this model, later forming bars/NSDs grow quicker than older ones. 

Earlier bar formation occurs in a less enriched galaxy, resulting in significantly lower pre-enrichment of the gas flowing onto the early NSD. Additionally, the overall mass is moderately larger as more stars can form over time.
These effects on the NSD mass are illustrated in Fig.~\ref{fig:gasmassbarformation}, showing the larger over-all stellar mass of an early-forming NSD. Conversely, the later-forming NSD grows more quickly, and has a slightly higher overall gas mass.

Next, Fig.~\ref{fig:metplanebarformation} examines the effects on chemical abundances. The final abundances in the onfall scenario are dominated by late-accreted gas, whose composition is consistent across our formation times (as it comes from nearly the same region of the galaxy), leading to similar final abundance profiles. However, a change in bar formation time leads to different inflowing gas composition at early times. This can be seen in the second row of Fig.~\ref{fig:metplanebarformation}, tracking the abundance of the gas feeding the NSD. Especially the iron abundance of the inflowing gas strongly depends on at which stage of galactic chemical evolution bar formation takes place. (See also Fig.~\ref{fig:GlobalIronAbundance} for the assumed galactic \hbox{[Fe/H]}\xspace evolution). This effectively stretched the trajectory in the metallicity plane to lower metallicities and higher $\alpha$-abundances, as seen in the lower left panels of Fig.~\ref{fig:metplanebarformation}.

While not strongly affecting the final abundance profiles in the onfall scenarios, the higher iron abundances of the inflowing gas at early times for late forming bars accumulates in the NSD centre and leads to a steeper \hbox{[Fe/H]}\xspace gradient. As the magnesium abundance of the inflowing gas is not strongly affected, the enhanced iron overabundance leads to a dip in \hbox{[Mg/Fe]$_\mathrm{OA}$}\xspace for late forming bars in the central NSD region.

\end{appendix}

\end{document}